\RequirePackage{ifpdf}
\RequirePackage{xcolor}
\documentclass[preprint, preprintnumbrs]{article}
\usepackage{jheppub}
\usepackage{xcolor}
\usepackage{amsmath}
\usepackage{epsfig}
\usepackage{caption}
\usepackage{subcaption}
\usepackage{soul}
\pdfoutput=1

\newcommand{\roughly}[1]{\mathrel{\raise.3ex\hbox{$#1$\kern-0.85em
\lower1ex\hbox{$\sim$}}}}
\newcommand{\lsim}{\roughly<}
\newcommand{\gsim}{\roughly>}

\def\nn{\nonumber}

\newcommand{\be}{\begin{equation}}
\newcommand{\bee}{\begin{equation}}
\newcommand{\ee}{\end{equation}}
\newcommand{\beea}{\begin{eqnarray}}
\newcommand{\eea}{\end{eqnarray}}
\newcommand{\bea}{\begin{eqnarray}}

\def\nott#1{\setbox0=\hbox{$#1$}                
   \dimen0=\wd0                                 
   \setbox1=\hbox{/} \dimen1=\wd1               
   \ifdim\dimen0>\dimen1                        
      \rlap{\hbox to \dimen0{\hfil/\hfil}}      
      #1                                        
   \else                                        
      \rlap{\hbox to \dimen1{\hfil$#1$\hfil}}   
      /                                         
   \fi}                                         %

\def\uxsl{\hbox{/\kern-.4000em$u$}}
\def\uxslsm{\hbox{\smaller/\kern-.5600em$u$}}
\def\pxpsl{\hbox{/\kern-.5000em$p$}}
\def\epssl{\hbox{/\kern-.5600em$\epsilon$}}
\def\delsl{\hbox{/\kern-.7000em$\nabla$}}
\def\lxpsl{\hbox{/\kern-.5600em$l$}}
\def\kxpsl{\hbox{/\kern-.5600em$k$}}
\def\qxpsl{\hbox{/\kern-.3900em$q$}}

\def\pref#1{(\ref{#1})}
\def\exd{{\rm d}}
\def\ol#1{{\overline{#1}}}

\def\cB{{\cal B}}
\def\cD{{\cal D}}
\def\cC{{\cal C}}

\def\cF{{\cal F}}
\def\cG{{\cal G}}
\def\cH{{\cal H}}

\def\cK{{\cal K}}
\def\cL{{\cal L}}

\def\cO{{\cal O}}

\def\cR{{\cal R}}

\def\cV{{\cal V}}
\def\cW{{\cal W}}

\def\cZ{{\cal Z}}

\def\mfa{{\mathfrak a}}
\def\mfb{{\mathfrak b}}

\def\mfB{{\mathfrak B}}

\def\mfG{{\mathfrak{G}}}

\def\ssA{{\scriptscriptstyle A}}

\def\ssC{{\scriptscriptstyle C}}
\def\ssD{{\scriptscriptstyle D}}

\def\ssM{{\scriptscriptstyle M}}
\def\ssN{{\scriptscriptstyle N}}

\def\ssW{{\scriptscriptstyle W}}
\def\ssX{{\scriptscriptstyle X}}
\def\ssY{{\scriptscriptstyle Y}}

\def\UV{{\scriptscriptstyle U\hbox{\kern-0.1em}V}}
\def\PPN{{\scriptscriptstyle P\hbox{\kern-0.1em}P\hbox{\kern-0.1em}N}}
\def\MN{{\scriptscriptstyle M\hbox{\kern-0.1em}N}}
\def\MNP{{\scriptscriptstyle M\hbox{\kern-0.1em}N\hbox{\kern-0.1em}P}}
\def\KK{{\scriptscriptstyle K\hbox{\kern-0.1em}K}}
\def\SM{{\scriptscriptstyle S\hbox{\kern-0.1em}M}}
\def\EH{{\scriptscriptstyle E\hbox{\kern-0.1em}H}}

\def\QCD{{\scriptscriptstyle Q\hbox{\kern-0.1em}C\hbox{\kern-0.1em}D}}
\def\IR{{\scriptscriptstyle I\hbox{\kern-0.1em}R}}
\def\TEV{{\scriptscriptstyle T\hbox{\kern-0.1em}E\hbox{\kern-0.1em}V}}
\def\aff{{a\hbox{\kern-0.1em}f\hbox{\kern-0.1em}f}}
\def\axion{{a}}
\def\baxion{{b}}

\title{UV and IR Effects in Axion Quality Control}

\author{C.P.~Burgess,${}^{1,2}$ Gongjun Choi${}^3$ and F.~Quevedo${}^{4}$ \\ 
{\it 
${}^1$ Department of Physics \& Astronomy, McMaster University\\ \qquad 1280 Main Street West, Hamilton ON, Canada.\\
${}^2$ Perimeter Institute for Theoretical Physics\\
\qquad 31 Caroline Street North, Waterloo ON, Canada.\\
${}^3$ CERN, Theoretical Physics Department, Gen\`eve 23, Switzerland.\\
${}^4$ DAMTP, University of Cambridge, Wilberforce Road,  Cambridge, CB3 0WA, UK.
}
}

\date{\today}

\abstract{Motivated by recent discussions and the absence of exact global symmetries in UV completions of gravity we re-examine the axion quality problem (and naturalness issues more generally) using antisymmetric Kalb-Ramond (KR) fields rather than their pseudoscalar duals, as suggested by string and higher dimensional theories. Two types of axions can be identified: a model independent $S$-type axion dual to a two form $B_{\mu\nu}$ in 4D and a $T$-type  axion coming directly as 4D scalar Kaluza-Klein (KK)  components of higher-dimensional tensor fields. For $T$-type axions our conclusions largely agree with earlier workers for the axion quality problem, but we also reconcile why $T$-type axions can couple to matter localized on 3-branes with Planck suppressed strength even when the axion decay constants are of order the KK scale. For $S$-type axions, we review the duality between form fields and massive scalars and show how duality impacts naturalness arguments about the UV sensitivity of the scalar potential. In particular UV contributions on the KR side suppress contributions on the scalar side by powers of $m/M$ with $m$ the axion mass and $M$ the UV scale. We re-examine how the axion quality problem is formulated on the dual side and compare to recent treatments. We study how axion quality is affected by the ubiquity of $p$-form gauge potentials (for both $p=2$ and $p=3$) in string vacua and identify two criteria that can potentially lead to a problem. We also show why most fields do not satisfy these criteria, but when they do the existence of multiple fields also provides mechanisms for resolving it. We conclude that the  quality problem is easily evaded.
 }

\preprint{CERN-TH-2022-176}

\begin{document}
\maketitle

\section{Introduction}

String theory giveth and string theory taketh away, at least where axions\footnote{We follow the string literature and broadly refer to any low-energy Goldstone boson enjoying a rigid compact shift symmetry as an `axion' (as opposed to the `dilatons' associated with rigid scaling symmetries), something that would be called an ALP (axion-like particle) by particle phenomenologists. Our later focus is on those Goldstone bosons whose symmetries have a QCD anomaly and so can take part in the strong-CP problem \cite{StrongCP, Weinberg:1977ma, Wilczek:1977pj} (which is what a particle physicist would usually mean by an `axion').} are concerned. On one hand axions are said to be ubiquitous in the spectrum of particles predicted around most string vacua \cite{StringUbiquity, stringaxions}. This observation motivates the study of their phenomenological consequences \cite{Axiverse}, with a particular focus of late on their possible role as a light form of dark matter \cite{AxionReviews}. 

On the other hand, string theory equally generally forbids\footnote{Although there are known ways out \cite{Burgess:2008ri} the conclusion is nonetheless broadly true and global symmetries tend to be both rare and approximate.} the existence of exact rigid (or global) symmetries \cite{NoGlobal}, in principle including the rigid shift symmetries on which low-energy axion properties are founded. For Goldstone bosons this breaking can keep them from being light, and can interfere with any mechanisms that rely on the survival of axions down to the low-energy theory. As applied to the QCD axion this has come to be known as the axion `quality' problem \cite{QualityProblem}.   

So which is it? Are axions as abundant as dirt or as diamonds in low-energy string vacua? The resolution (which has long been known) is that there is a sense they are both. The absence of global symmetries really does mean that one never really directly finds scalar axions $a$ with shift symmetries in string vacua. Instead these scalars arise indirectly as Kaluza-Klein (KK) modes from fields that not themselves scalars; commonly arising\footnote{4D axions can also arise as KK modes from other types of extra-dimensional fields, but we focus on the Kalb-Ramond field because it allows a unified treatment of two different types of 4D axion.} as components of 2-form Kalb-Ramond gauge fields \cite{KRGauge}, $B = \frac12 \, B_{\MN} \, \exd z^\ssM \wedge \exd z^\ssN$, subject to the gauge symmetries $B \to \exd \lambda$ for some arbitrary field $\lambda_\ssM(x)$. Fields like $B_{\MN}$ arise so frequently in string vacua because they are related to other fields (notably the metric) by supersymmetry in higher dimensions. 

\subsection{Types of UV axion pedigree}

Low-energy scalars typically emerge in the 4D effective theory from such fields in one of two ways:
\begin{itemize}
\item {\bf $T$-type axions:} $b(x)$ are specific cases of Kaluza-Klein (KK) modes arising when dimensionally reducing the extra-dimensional components $B_{mn}(x,y) = b(x) \, \omega_{mn}(y)$, where $x^\mu$ denote the observed 4 dimensions, $y^m$ are extra-dimensional coordinates and $\omega_{mn}(y)$ is a harmonic 2-form field within the extra dimensions. 
\item  {\bf $S$-type axions:} $\axion(x)$ arise directly as the 4-dimensional components $B_{\mu\nu}(x,y) = b_{\mu\nu}(x) \,\omega(y)$,
which in four dimensions are known to be dual to scalar fields with shift symmetries \cite{Savit:1979ny} through relations of the form $\partial^\mu \axion \propto \epsilon^{\mu\nu\lambda\rho} \partial_{\nu}B_{\lambda\rho}$ (much more about which below). Here $\omega(y)$ is a harmonic 0-form field -- typically a $y$-independent constant that can depend on extra-dimensional moduli.
\end{itemize}
This UV provenance is of course relevant to the axion quality problem, which is in essence an issue of UV sensitivity. One of our goals with this paper is to explore the ways that it helps, for both $T$- and $S$-type axions. Some of our conclusions are similar to earlier discussions of this issue \cite{Kallosh:1995hi, DualStrongCP}, in particular that the problem gets rephrased in dual form (for $S$-type axions) in terms of the existence of multiple 3-form gauge potentials. 

Since these issues have recently been revisited anew \cite{Sakhelashvili:2021eid, Dvali:2022fdv} we clarify what properties these fields must have to actually cause a quality problem and use this to argue why gravitational examples specifically (and the great abundance of such potentials in string vacua more generally) need not pose a problem in themselves. The dual formulation also suggests how the presence of multiple axions (as is common in string theory) can help alleviate the quality problem. The upshot is that the UV can, but need not, cause a quality problem. Whether or not it does cannot be decided purely at low energies because it depends on what happens in the UV.\footnote{The same is is also true of other naturalness problems; they arise because of strong dependence on physical masses for states that actually appear in the UV theory and not a dependence on cutoffs, as is sometimes mistakenly asserted (for a summary of these issues see {\it e.g.}~\cite{Burgess:2013ara}).}

But our discussion has implications that apply more broadly than just to the quality problem for the QCD axion. Along the way we identify more generally how dimensional `naturalness' arguments for the scalar potential give very different estimates depending on whether they are made directly for the scalar or are first done for its dual and then mapped to the scalar using duality. In particular terms involving $n$ powers of the canonically normalized scalar arise additionally suppressed by powers of $(m/M)^n$ where $M$ is the UV scale and $m$ is the axion mass (an observation also made in the past for inflationary models \cite{NaturalnessForms}). 

We find a number of other ways that axion properties suggested by string-motivated extra-dimensional physics can be informative. For instance we describe a simple model for which $T$-type axions have physical axion-matter couplings $g_{\aff}$ that are dramatically smaller than the naive value $1/f$ read off from the axion kinetic term. In the example given (motivated by the models of \cite{YogaDE}) $g_{\aff}$ is order $1/M_p$ despite $f$ being an ordinary particle-physics scale. Decoupling these scales from one another could have practical implications for axion phenomenology.  

We show why the same hierarchy does not arise in these models for $S$-type axions and we clarify why not. Physical couplings of $S$-type axions really are of order $1/f$ and we identify which interactions in the UV completion are responsible for the breakdown of the $E/f$ expansion at energies $E \gsim f$. $S$-type axions illustrate how the scalar and dual representations can provide instances of weak/strong coupling duality, for which both the scalar and the dual cannot be within the weakly coupled regime. In the extra-dimensional example studied it is the Kalb-Ramond formulation that is weakly coupled. This could also have phenomenological implications to the extent that an axion that is dual to a weakly coupled system is unlikely to be well-described by the semiclassical methods that are universally used when exploring its physical implications.

Some of these observations imply that the use of the scalar (rather than Kalb-Ramond) variable can be misleading in some circumstances. This can seem surprising at first sight because the duality between axions and Kalb-Ramond fields is in essence a field redefinition and so scalar and dual formulations should be completely equivalent; it shouldn't matter that string theory hands you Kalb-Ramond fields if scalar axions are equivalent and are much simpler to work with. Why should one care that a more complicated framework exists if it only obscures implications drawn using more transparent methods?  We argue here that phenomena like weak/strong coupling duality  are special cases of Weinberg's Third Law of Progress in Theoretical Physics \cite{Weinberg:1981qq}: {\it You can use any degrees of freedom you like to describe a physical system, but if you use the wrong ones you'll be sorry.}

\subsection{Non-propagating low-energy forms}

These duality arguments touch on a related rich vein of physics with broader significance: the importance of keeping non-propagating entities like auxiliary and/or topological fields when formulating Wilsonian effective theories. These are fields that can be integrated out without changing the types of particles that propagate, and so it is tempting to think one should do so once and for all and simply ignore them thereafter. However such fields bring to the low-energy effective theory information about how its UV completion responds, {\it e.g.}~to environments with nontrivial topology. They arise in concrete situations (such as in EFTs for 3-dimensional Quantum Hall systems, where the presence of emergent non-propagating gauge fields is essential for capturing the fractional quantization of Hall plateaux and the unusual charge and statistics of some excitations \cite{QHE, EFTBook}). 

Evidence is building that a similar role is played more widely by 3-form gauge potentials in four spacetime dimensions, $C := \frac16 \, C_{\mu\nu\lambda} \, \exd x^\mu \exd x^\nu \exd x^\lambda$ subject to the gauge freedom $C \to C + \exd \Lambda$ where $\Lambda_{\mu\nu}(x) = - \Lambda_{\nu\mu}(x)$ is an arbitrary 2-form field. These are known to bring to the low-energy 4D effective theory topological information coming from integrated out extra dimensions \cite{Bousso:2000xa, Burgess:2015lda}, and more generally provide the origin for the auxiliary fields that appear in the 4D supergravities that are the low-energy limits of string vacua \cite{Bielleman:2015ina, Herraez:2018vae}. They appear in the QCD quality story because they can give masses to Kalb-Ramond fields \cite{Quevedo:1996uu} through a Higgs mechanism that is dual to more mundane methods of axion mass generation. Because the field strength $H = \exd C$ often appears in the action with a definite sign (often as a square), its presence can alter the implications of naturalness arguments for the scalar potential  \cite{Burgess:2021juk}. Indeed such terms provide the 4D understanding of why 6D SLED models \cite{SLED} can in some circumstances suppress the 4D vacuum energy, but also why they struggle to do so enough to solve the cosmological constant problem \cite{Burgess:2015gba, Burgess:2015lda, Burgess:2015kda, Niedermann:2015via}. Their interplay with accidental scaling symmetries lies behind a recent attempt to find a dynamical relaxation mechanism for vacuum energies in four dimensions \cite{YogaDE}. 

In what follows we build our case for the above story using concrete examples. We first, in \S\ref{sec:DualAxions}, briefly review the duality construction -- in particular its extension to massive axions \cite{Quevedo:1996uu}, which provides a way to think about scalar masses arising through a Higgs mechanism. \S\ref{sec:NaturalDual} then briefly reviews and clarifies its use to dualize the axion solution to the strong-CP problem \cite{DualStrongCP}, highlighting in particular how the quality problem gets rephrased in the dual language and how the apparent UV sensitivity of terms in the EFT differs between the axion formulation and its dual. Finally \S\ref{sec:SemiClass} provides a concrete extra-dimensional example -- inspired by a UV completion of \cite{YogaDE} -- that illustrates both how axion/Kalb-Ramond duality can map weak to strong couplings, and how enormous hierarchies can arise with $g_{\aff}, g_{a\gamma\gamma} \sim 1/M_p$ even with $f$ as low as eV scales.   

\section{Axions and duality}
\label{sec:DualAxions}

We start with a review of why 2-form gauge potentials like $B_{\mu\nu}$ are dual \cite{Savit:1979ny, Buscher:1987qj} to scalar fields, both in the standard shift-symmetric massless case and for massive scalars, following the discussion of \cite{Quevedo:1996uu} (which in turn generalizes earlier arguments \cite{JuliaToulouse} aimed at describing particle/vortex duality in Kosterlitz-Thouless transitions \cite{KTTransitions}).

\subsection{Axion/2-form duality}

Consider the following path integral
\be
  \Xi[J] =  \int \cD B \; e^{iS_1[B]}
\ee
where $S_1 = \int \exd^4 x \; \cL_1$ with $\cL_1$ chosen (at least to start) to be 
\be \label{KRForm1}
   \cL_1(B) =  - \frac{\cZ}{2\cdot 3!} \, G_{\mu\nu\lambda} G^{\mu\nu\lambda}  -   \frac{1}{3!} \, \epsilon^{\mu\nu\lambda\rho} G_{\mu\nu\lambda} J_\rho   \,,
\ee
with $G = \exd B$ the exterior derivative of a 2-form field $B_{\mu\nu}$ and $\cZ$ and $J_\rho$ possibly depending on other fields (collectively denoted $\psi$). $B = \frac12 \, B_{\mu\nu} \, \exd x^\mu \wedge \exd x^\nu$ is only defined up to the gauge redundancy $B \to B + \exd  \lambda$ for an arbitrary 1-form $\lambda = \lambda_\mu \, \exd x^\mu$. 

The duality starts by trading the integration over $B_{\mu\nu}$ for an integral over $G_{\mu\nu\lambda}$ subject to a constraint that imposes the Bianchi identity $\exd G = 0$. These are equivalent because the Bianchi identity is sufficient to guarantee the local existence of a field $B_{\mu\nu}$ with $G = \exd B$. The constraint is imposed by integrating over a scalar Lagrange-multiplier field $\axion$, and so writing
\be
  \Xi[J] =  \int \cD G  \, \cD \axion \; e^{iS_0}
\ee
where $S_0 = \int \exd^4 x \; \cL_0$ with
\be \label{MasterDual}
   \cL_0(G,\axion) =  - \frac{\cZ}{2\cdot 3!} \, G_{\mu\nu\lambda} G^{\mu\nu\lambda} - \frac{1}{3!} \axion \, \epsilon^{\mu\nu\lambda\rho} \partial_\mu G_{\nu\lambda\rho}  - \frac{1}{3!} \, \epsilon^{\mu\nu\lambda\rho} G_{\mu\nu\lambda} J_\rho \,,
\ee
Integrating out $\axion$ imposes the Bianchi identity $\exd G = 0$ and allows the integral over $G$ to be replaced with the integral over $B$, leading back to \pref{KRForm1}. 

The dual version is obtained from \pref{MasterDual} by instead integrating out $G_{\mu\nu\lambda}$ so that $\axion$ is the remaining field. The result inherits a shift symmetry $\axion \to \axion \;+ $ constant because $\cL_0$ transforms into a total derivative. The $G$ integration is gaussian, whose saddle point is $G_{\mu\nu\lambda} = \cG_{\mu\nu\lambda}$ where
\be \label{GSaddle}
    \cG_{\mu\nu\lambda} = - \cZ^{-1} \epsilon_{\mu\nu\lambda\rho} \Bigl( \partial^\rho \axion + J^\rho \Bigr) \,,
\ee
and so the integration gives the new lagrangian density
\be \label{Dualmassless}
   \cL_2(\axion) 
   =  - \frac{1}{2\cZ} (\partial_\mu \axion + J_\mu)( \partial^\mu \axion +  J^\mu) \,.
\ee
If $\cZ = 1$ then $\axion$ is a canonically normalized massless scalar derivatively coupled to the same local current $J_\mu$ as in the original formulation. Because \pref{KRForm1} and \pref{Dualmassless} are both obtained from \pref{MasterDual} they must describe equivalent physics. Although the implied field redefinition from $B_{\mu\nu}$ to $\axion$ is in principle nonlocal the physics on both sides is nonetheless local because this is true of the relation between the field strengths given in \pref{GSaddle}.  

\subsubsection*{Significance of $\cZ \leftrightarrow \cZ^{-1}$}

In reality the above gaussian action is always supplemented by other non-gaussian interactions $\cL_{\rm int}$ within a low-energy Wilsonian effective field theory (EFT). To the extent that both $B_{\mu\nu}$ and $\axion$ are derivatively coupled perturbative semiclassical methods in the presence of nongaussian terms like $(G_{\mu\nu\lambda}G^{\mu\nu\lambda})^2 \in \cL_{\rm int}$ are ultimately justified by a low-energy derivative expansion that applies equally well on both sides of a duality relationship because relationships like \pref{GSaddle} involve equal numbers of derivatives on both sides. 

The inversion of $\cZ \to \cZ^{-1}$ as one passes from \pref{KRForm1} to \pref{Dualmassless} is a noteworthy feature of duality. When $\cZ \gg 1$ this implies 2-point correlators of $G_{\mu\nu\lambda}$ are order $\cZ^{-1}$ in size while those of $\partial_\mu\axion$ are instead order $\cZ$. The significance of the change $\cZ \to \cZ^{-1}$ depends on whether or not $B_{\mu\nu}$ and $\axion$ can be freely rescaled to remove $\cZ$ by going to canonically normalized variables. If this is so then $\cZ$ in any case drops out of observables. For instance, when $J_\mu \neq 0$ this rescaling shows that $\Xi$ is really only a function of $\widetilde J_\mu := \cZ^{-1/2} J_\mu$ rather than depending on $\cZ$ and $J_\mu$ separately. Although $\cZ \leftrightarrow \cZ^{-1}$ is sometimes called weak/strong coupling duality, $\Xi[\widetilde J]$ is the same on both sides of the duality and so both sides agree on its functional dependence if expanded order-by-order in powers of $\cZ^{-1}$ (say). 

One situation where this kind of rescaling is not possible is when $\cZ$ depends on other fields and the target-space metric in field space is not flat. Another case where physics can depend explicitly on $\cZ$ is when the field $B_{\mu\nu}$ or $\axion$ is quantized\footnote{This is generic the case in string theory for which the symmetries associated to antisymmetric tensors and axions are compact (meaning there always exist magnetic-like branes). For a general discussion see \cite{Banks:2010zn}.}, perhaps satisfying a boundary condition like $\oint_\ssW \exd x^\mu \partial_\mu \axion = 2\pi n f$ for some curve $\cW$, integer $n$ and mass scale $f$, or perhaps $\oint_\ssC B = 2\pi \tilde n f^{-1}$ for some 2-cycle $C$ and possibly different integer $\tilde n$ and mass scale $\tilde f$. In these situations physical results can depend on $\cZ$ ({\it i.e.}~on $f$ and/or $\tilde f$) and $J_\mu$ separately, and the relation $\cZ \to \cZ^{-1}$ can carry physical significance.

\subsection{A Higgs mechanism for scalar masses}

Although the above makes the shift symmetry (and so also masslessness) of $\axion$ seem automatic, we next summarize how duality extends to massive scalars, following \cite{Quevedo:1996uu}. A scalar potential is achieved in the dual framing through a Higgs mechanism in which the field $B_{\mu\nu}$ `eats' (or is eaten by) a non-propagating gauge potential\footnote{Known string vacua can also contain a large number of these 3-form gauge potentials.} $C_{\mu\nu\lambda}$. Because $C_{\mu\nu\lambda}$ does not propagate this meal does not change the number of propagating degrees of freedom.
 
To this end consider the following gaussian path integral
\be
  \Xi[J] = \int \cD C \, \cD B  \; e^{iS_1}
\ee
where $S_1 = \int \exd^4 x \; \cL_1$ and
\bea
   \cL_1(C,B) &=& - \frac{1}{2\cdot 4!} H_{\mu\nu\lambda\rho} H^{\mu\nu\lambda\rho} - \frac{1}{2\cdot 3!} (G_{\mu\nu\lambda} + mC_{\mu\nu\lambda})( G^{\mu\nu\lambda} + m C^{\mu\nu\lambda}) \nn\\
   &&\qquad \qquad   - \frac{1}{3!} \, \epsilon^{\mu\nu\lambda\rho} ( G_{\mu\nu\lambda} + m C_{\mu\nu\lambda})  J_\rho  \,.
\eea
Here $C_{\mu\nu\lambda}$ is a 3-form gauge potential with field strength $H = \exd C$ and $B_{\mu\nu}$ is a 2-form gauge potential with $G = \exd B$ while $m$ is a parameter with dimension mass. 

This lagrangian has the gauge symmetry $C \to C + \exd \Lambda$ and $B \to B - m\, \Lambda$ for an arbitrary 2-form $\Lambda$. So when $m \neq 0$ we can choose a gauge $B = 0$. The field equation for $C$ that follows from this action then is
\be
  D_\mu H^{\mu\nu\lambda\rho} + m^2 C^{\nu\lambda\rho} + m\, \epsilon^{\nu\lambda\rho\mu} J_\mu = 0 \,.
\ee
This describes a single spin state propagating with mass $m$ once all the gauge symmetries are used, as can be seen by counting the massless states from which it is built. (In 4D $B_{\mu\nu}$ is shown above to be equivalent to a massless scalar and $C_{\mu\nu\lambda}$ contains no propagating degrees of freedom at all because one can always write $H_{\mu\nu\lambda\rho} = h \, \epsilon_{\mu\nu\lambda\rho}$ with field equation $\partial_\mu H^{\mu\nu\lambda\rho} = 0$ in the massless limit, which implies $h$ is a constant and so does not propagate.)

The dual should therefore be a massive scalar and this can be verified by trading the integral over $B$ for an integral over $G$ and introducing (as before) a lagrange multiplier $\axion$ to impose the Bianchi identity\footnote{One can equivalently omit the $mC_{\mu\nu\lambda}$ terms everywhere and instead impose the modified Bianchi identity $\exd G = m H$.} $\exd G = 0$, leading to the lagrangian density
\bea
   \cL_0(C,G,\axion) &=& - \frac{1}{2\cdot 4!} H_{\mu\nu\lambda\rho} H^{\mu\nu\lambda\rho} - \frac{1}{2\cdot 3!} (G_{\mu\nu\lambda} + mC_{\mu\nu\lambda})( G^{\mu\nu\lambda} + m C^{\mu\nu\lambda}) \nn\\
   &&\qquad \qquad - \frac{1}{3!}\, \axion \, \epsilon^{\mu\nu\lambda\rho} \partial_\mu G_{\nu\lambda\rho}  - \frac{1}{3!} \, \epsilon^{\mu\nu\lambda\rho} ( G_{\mu\nu\lambda} + m C_{\mu\nu\lambda})  J_\rho  \,.
\eea
Integrating out $\axion$ returns us to the above formulation, but instead performing the integration over $G$ leads to the saddle point 
\be
    \cG_{\mu\nu\lambda} = - m C_{\mu\nu\lambda} - \epsilon_{\mu\nu\lambda\rho} \Bigl( \partial^\rho \axion + J^\rho \Bigr) \,,
\ee
and so to the lagrangian 
\be
   \cL_2(C,\axion) 
   = - \frac{1}{2\cdot 4!} H_{\mu\nu\lambda\rho} H^{\mu\nu\lambda\rho}  - \frac{m}{4!}\,  \axion \, \epsilon^{\mu\nu\lambda\rho} H_{\mu\nu\lambda\rho}- \frac12 \partial_\mu \axion \, \partial^\mu \axion - J^\mu \partial_\mu \axion - \frac12 J_\mu J^\mu  \,.
\ee

Next we perform the integral over $C_{\mu\nu\lambda}$, and this is equivalent to simply performing the gaussian integral over $H_{\mu\nu\lambda\rho}$ because the integrability condition for writing $H = \exd C$ is $\exd H = 0$ which is always true (in 4D). The saddle point for the $H$ integral occurs for $H_{\mu\nu\lambda\rho} = \cH_{\mu\nu\lambda\rho}$ where
\be \label{HSaddle}
   \cH_{\mu\nu\lambda\rho} = - m \, \axion \, \epsilon_{\mu\nu\lambda\rho}
\ee
and so leads to the scalar lagrangian
\be
   \cL_2(\axion)
  = - \frac12 (\partial \axion)^2 - \frac{m^2}{2} \axion^2 - J^\mu \partial_\mu \axion - \frac12 J_\mu J^\mu  \,.
\ee
This is the expected massive scalar.

\subsubsection{Scalar potential}
\label{ssec:SPot}

For future reference notice that it is only this last step that would differ if we'd had higher-dimension terms like $\delta \cL = W(X)$ in the lagrangian with $X = \frac{1}{4!} \epsilon^{\mu\nu\lambda\rho}H_{\mu\nu\lambda\rho}$ and so $X^2 = - \frac{1}{4!} H_{\mu\nu\lambda\rho}H^{\mu\nu\lambda\rho}$ and so on. The above discussion is the special case $W = \frac12 X^2$ but one could entertain, for example,
\be \label{Wquart}
   W = c_1 M^2 X + \frac12 \, X^2 + \frac{2c_3}{3M^2} \, X^3 + \frac{c_4}{4M^4} \, X^4 + \cdots
\ee
where the coefficients $c_i$ are dimensionless and $M$ is a UV scale inserted everywhere on dimensional grounds (with $H_{\mu\nu\lambda\rho}$ canonically normalized). 

For non-quadratic $W$ the integral over $H$ is no longer gaussian, but we can proceed assuming a semiclassical saddle-point approximation is valid, in which case the saddle point \pref{HSaddle} is modified to
\be \label{alphavsWX}
    \left( \frac{\partial W}{\partial X} \right)_{H = \cH} = m\, \axion \,,
\ee
which agrees with \pref{HSaddle} when $W =  \frac12 X^2$. For example, for the choice \pref{Wquart} this becomes
\be
  c_1 M^2 + X\left( 1 + \frac{2c_3}{M^2} X + \frac{c_4}{M^4} X^2 + \cdots \right) \simeq m \, \axion
\ee
and so
\be
   X \simeq m\, \axion - c_1 M^2 - \frac{2c_3}{M^2}\left( m\, \axion - c_1 M^2 \right)^2 
    + \cO\left[\left( m\, \axion - c_1 M^2 \right)^3/M^4\right] \,.
\ee

Once used in the lagrangian this shows how non-quadratic pieces of $W$ map over to non-quadratic contributions to the scalar potential for $\axion$ in the dual lagrangian $\cL_2$.  In particular the axion potential becomes
\be
  V(\axion) = - W(X) + m \axion X = \frac12 \left(m \axion - c_1 M^2 \right)^2 - \frac{2c_3}{3M^2}  \left(m\axion - c_1 M^2 \right)^3 + \cdots \,. 
\ee
Two features are noteworthy about this potential:
\begin{itemize}
\item
First, notice it shares the usual Legendre property
\be \label{dVdalphavsX}
  \frac{\partial V}{\partial \axion} = m X + \left( - \frac{\partial W}{\partial X} + m \axion \right) \frac{\partial X}{\partial \axion} = mX\,,
\ee
where the last equality uses \pref{alphavsWX}. Even if new non-quadratic terms introduce new stationary points for $V(\axion)$ (or shifts the positions of old ones) eq.~\pref{dVdalphavsX} ensures $X = 0$ for {\it all} of them. 
\item 
Second, once $\axion$ is shifted so that the minimum is at $\axion = 0$ the potential depends on $m$ and $\axion$ only through the combination $m \axion$. Consequently, a term proportional to $\axion^n$ comes suppressed by a power of $(m/M)^n$ relative to what would naively be expected on dimensional grounds for $V(\axion)$. This is how the dual theory reproduces the same $M$-dependence as found for higher powers of $H_{\mu\nu\lambda\rho}$ given that $\axion$ has canonical dimension mass while $H$ has dimension (mass)${}^2$. This shows how a dimensional assessment of how UV scales appear in the low-energy theory can care about the existence of a dual formulation. 
\end{itemize}

\section{Naturalness issues for dual systems}
\label{sec:NaturalDual}

This section examines how naturalness arguments look for $T$- and $S$-type axions, and for $S$-type axions how they depend on which side of the duality relation they are made. We do so using the axion quality problem as a representative example.

\subsection{QCD and the dual PQ mechanism}
\label{ssec:DualQCDPQ}

To this end we extend the above reasoning to the main event: QCD and the $\theta$-term. The idea is to dualize the coupling of the axion to QCD to see how the strong-CP problem gets formulated, along the general lines of \cite{DualStrongCP}. We then ask how UV physics might complicate the story in the dual theory. Consider then adding a gauge potential $A_\mu$ (with field strength $F_{\mu\nu}$) to represent the QCD gauge sector\footnote{We do not write quarks explicitly but flag the few places where their implicit presence affects what is written.} and this time consider the path integral
\be
  \Xi[J] = \int  \cD G \, \cD A \, \cD \axion \; e^{iS_0}
\ee
where $S_0 = \int \exd^4 x \; \cL_0$ and
\bea \label{QCDAxialDual}
   \cL_0(G,A,\axion) &=&  - \frac{1}{2\cdot 3!} G_{\mu\nu\lambda} G^{\mu\nu\lambda}  -\frac{ \axion}{3!} \, \epsilon^{\mu\nu\lambda\rho} \left(\partial_\mu G_{\nu\lambda\rho} - \frac14\, \Omega_{\mu\nu\lambda\rho} \right)  - \frac{1}{3!} \, \epsilon^{\mu\nu\lambda\rho} G_{\mu\nu\lambda}  J_\rho  \nn\\
   && \qquad\qquad  - \frac14 F_{\mu\nu} F^{\mu\nu} - \frac{\theta}{2} \, \epsilon^{\mu\nu\lambda\rho} F_{\mu\nu}F_{\lambda\rho}  \,.
\eea
We suppress both gauge-group indices and traces over them to avoid notational clutter. $F_{\mu\nu}$ is the field strength for the gauge potential $A_\mu$ but $G_{\mu\nu\lambda}$ is an arbitrary 3-form until the integral over $\axion$ is performed. 

Integrating out $\axion$ imposes the Bianchi identity $\exd G = \Omega$ where $\Omega$ is a gauge-invariant quantity built from the gauge field that on grounds of consistency must satisfy $\exd \Omega = 0$, for which we take
\be \label{OmegaDefFF}
    \frac{1}{12}\,\epsilon^{\mu\nu\lambda\rho}\Omega_{\mu\nu\lambda\rho} = \frac{1}{f} \, \epsilon^{\mu\nu\lambda\rho} F_{\mu\nu} F_{\lambda\rho}
\ee
The mass scale $f$ is here required on dimensional grounds. Doing this allows the $G$ integral to be traded for one over $B$ as before and gives the lagrangian 
\be \label{QCDAxialDualNoAlpha}
   \cL_1(B,A) =  - \frac{1}{2\cdot 3!} G_{\mu\nu\lambda} G^{\mu\nu\lambda}    - \frac{1}{3!} \, \epsilon^{\mu\nu\lambda\rho}  G_{\mu\nu\lambda}  J_\rho    - \frac14 F_{\mu\nu} F^{\mu\nu}- \frac{\theta}{2} \, \epsilon^{\mu\nu\lambda\rho} F_{\mu\nu}F_{\lambda\rho}    \,.
\ee
where $G = \exd B + S$ where $\exd\Omega = 0$ implies there locally exists an $S_{\mu\nu\lambda}$ -- the Chern-Simons 3-form -- that satisfies $\Omega = \exd S$. 

The dual formulation instead integrates out $G$ and leaves $\axion$ as the dual variable.  Integrating out $G$ leads to the lagrangian density
\bea
   \cL_2(A,\axion) &=& - \frac12 (\partial \axion)^2 - J^\mu \partial_\mu \axion - \frac12 J_\mu J^\mu   + \frac{ \axion}{4!} \, \epsilon^{\mu\nu\lambda\rho}  \Omega_{\mu\nu\lambda\rho}   - \frac14 F_{\mu\nu} F^{\mu\nu}- \frac{\theta}{2} \, \epsilon^{\mu\nu\lambda\rho} F_{\mu\nu}F_{\lambda\rho}\nn\\
   &=& - \frac12 (\partial \axion)^2    - J^\mu \partial_\mu \axion - \frac12 J_\mu J^\mu    - \frac14 F_{\mu\nu} F^{\mu\nu} + \frac12\left( \frac{ \axion}{f}-\theta \right) \, \epsilon^{\mu\nu\lambda\rho} F_{\mu\nu}F_{\lambda\rho} \,.
\eea
This shows that the standard axion-gauge coupling is the dual of the 2-form/QCD coupling given in $\cL_1$ and that $f$ can be interpreted as its decay constant.

\subsubsection*{Below the QCD scale}

In the standard axion-QCD story integrating out QCD leaves a residual axion potential due its anomalous coupling to $F \wedge F$. This minimum is argued to be minimized where $\axion = \ol\theta \, f$ (where $\ol\theta$ is the usual combination of $\theta$ and phases in the quark mass matrices) which ensures that the CP-odd contribution turns off. We seek to express how physics below the QCD scale works in the dual language involving $B_{\mu\nu}$. 

Below $\Lambda_\QCD$ the gauge degrees of freedom are integrated out, naively leaving only hadrons coupled to $B_{\mu\nu}$. The key thought is that this is not quite right: the QCD EFT below $\Lambda_\QCD$ contains a path integral over low-energy hadrons {\it and} an integration over a low-energy field $C_{\mu\nu\lambda}$, whose emergent presence the strongly coupled vacuum of QCD makes mandatory. The field $C_{\mu\nu\lambda} \propto \langle S_{\mu\nu\lambda} \rangle$ is the low-energy counterpart of the Chern-Simons field appearing in the topological susceptibility \cite{Luscher:1978rn} above the QCD scale, where $\exd S =  F \wedge F$. 

Having this field in the low-energy theory below the QCD scale does not affect the existence of a gap or the spectrum of the known hadrons because $C_{\mu\nu\lambda}$ does not propagate. It is an auxiliary field that is required in order for the low-energy theory to capture properly the response of QCD to any topology in its environment. Similar fields are known to arise in this way in other concrete systems like the EFTs describing Quantum Hall systems \cite{QHE, EFTBook}. This 3-form potential differs from many of the others that often arise in string vacua because it arises from the IR properties of QCD rather than from the physics of UV compactification.

On dimensional grounds we write $H = \exd C$ with
\be \label{HvsFF}
  \frac{1}{12} \tilde\Lambda_\QCD^2 \epsilon^{\mu\nu\lambda\rho} H_{\mu\nu\lambda\rho} = \epsilon^{\mu\nu\lambda\rho} \langle F_{\mu\nu} F_{\lambda\rho} \rangle \,,
\ee
where $\tilde \Lambda_\QCD$ denotes a parameter of order the QCD scale that ensures that $H$ has canonical dimension (mass)${}^2$. The lagrangian \pref{QCDAxialDualNoAlpha} above the QCD scale is then replaced with its low-energy counterpart
\be \label{BelowQCD}
   \cL_1(C,B ) = - \frac{1}{2\cdot 3!} G_{\mu\nu\lambda}  G^{\mu\nu\lambda}  - \frac{1}{3!} \, \epsilon^{\mu\nu\lambda\rho}  G_{\mu\nu\lambda}  J_\rho - \frac{\ol\theta}{4!} \, \tilde \Lambda_\QCD^2 \epsilon^{\mu\nu\lambda\rho} H_{\mu\nu\lambda\rho}       - \frac{1}{2\cdot 4!} H_{\mu\nu\lambda\rho} H^{\mu\nu\lambda\rho}  + \cdots\,,
\ee
where the explicit term proportional to $\theta X$ combines with quark mass phases -- that also enter as terms linear in $X$, as in the $c_1$ term of \pref{Wquart} -- to produce $\ol\theta X$. The ellipses in \pref{BelowQCD} are at least cubic in $X$ (or involve derivatives of $X$). 

Combining eq.~\pref{OmegaDefFF} (and the discussion just above it) with \pref{HvsFF} implies
\be
  \exd G = \langle \Omega \rangle = \frac{\tilde \Lambda_\QCD^2}{f} \, H \,,
\ee
and so comparing this to $\exd G = m H$ (as would follow from $G = \exd B + m C$) allows us to read off the mass relation $m = \tilde \Lambda_\QCD^2/f$. We see that the $mC$ term captures the expectation value $\langle S \rangle/f$ of the Chern-Simons term in the UV theory above the QCD scale if $m$ scales with $f$ in the same way that the usual axion mass depends on its decay constant.

We expect the low-energy presence of such a 4-form field $H$ to give $B$ a nonzero mass, as we check by introducing the lagrange multiplier $\axion$ in the usual way and integrating out $G$. This leads to the result
\be  \label{QCDAxialDualx2}
   \cL_2(C, \axion) = - \frac12 (\partial \axion)^2 - J^\mu \partial_\mu \axion - \frac12 J_\mu J^\mu  +   \frac{1}{4!} (m\axion - \ol\theta \tilde \Lambda_\QCD^2 )\epsilon^{\mu\nu\lambda\rho} H_{\mu\nu\lambda\rho}  - \frac{1}{2\cdot 4!} H_{\mu\nu\lambda\rho} H^{\mu\nu\lambda\rho}  + \cdots\,.
\ee

Integrating out $H$ leads to the saddle point $H_{\mu\nu\lambda\rho} = \cH_{\mu\nu\lambda\rho}$ with
\be \label{lowestZvsalpha}
   \cH_{\mu\nu\lambda\rho} = \left(m\axion - \ol\theta \tilde \Lambda_\QCD^2 \right) \,\epsilon_{\mu\nu\lambda\rho} \,, 
\ee
and so gives the axion lagrangian
\be
   \cL_2(  \axion)
   = - \frac12 (\partial \axion)^2    - J^\mu \partial_\mu \axion - \frac12 J_\mu J^\mu  - \frac12 \left( m \axion -\ol\theta \tilde \Lambda_\QCD^2 \right)^2  \,,
\ee
showing that the minimum indeed occurs where $\axion = \ol\theta \tilde \Lambda_\QCD^2/m = \ol\theta f$, which turns off the CP-violating term of \pref{QCDAxialDualx2}. 

In general integrating out the UV QCD sector also generates more complicated low-energy interactions involving $C$, such as the function $W(X)$ of $X = \frac{1}{4!} \epsilon^{\mu\nu\lambda\rho} H_{\mu\nu\lambda\rho}$. As above, such terms semiclassically change the saddle point to 
\be \label{alphavsWX2}
    \left( \frac{\partial W}{\partial X} \right)_{H = \cH} = m\, \axion - \ol\theta \tilde \Lambda_\QCD^2 \,,
\ee
and so leads to the axion potential
\be
  V(\axion) = - W(X) + (m \axion - \ol\theta \tilde \Lambda_\QCD^2) X  \,.
\ee
This satisfies
\be \label{dVdalphavsX2}
  \frac{\partial V}{\partial \axion} = m X + \left( - \frac{\partial W}{\partial X} + m \axion - \ol\theta \tilde \Lambda_\QCD^2 \right) \frac{\partial X}{\partial \axion} = mX\,,
\ee
and so again ensures that $X = 0$ at {\it any} of the stationary points of $V$. We see that the presence of interactions like $W(X)$ show that $V$ is minimized at $m \axion = \ol\theta \tilde \Lambda_\QCD^2$ if $\partial W/\partial X$ vanishes when $X=0$.

\subsection{The Quality Problem}

We now have the tools required to explore UV sensitivity and the axion quality problem. We start by restating the original formulaton of the quality problem and then how it is rephrased in 2-form language for both $T$-type (this section) and $S$-type (next section) axions. 

The axion quality problem asks two related questions \cite{QualityProblem}: 
\begin{enumerate}
    \item Do corrections to the QCD axion potential change its minimum in a way that preserves a sufficiently small effective vacuum angle: $\bar{\theta}_{\rm eff}\lesssim10^{-10}$?
    \item Do corrections to the QCD axion potential change the usual expression for the axion mass (that assumes it is dominantly generated by the  `IR-dominated'  QCD instanton with size $\rho\sim\Lambda_{\QCD}^{-1}$)?
\end{enumerate}
The first of these essentially asks if the QCD axion remains a good solution to the strong CP problem when perturbed by new physics, whereas the second asks the same of our understanding of axion mass. The axion mass question can apply more generally to ALPs as well, whereas the first one is specific to the QCD axion.

Any UV completion must decide what happens at energies above the axion decay constant $f$ above which the low-energy expansion in powers of $E/f$ breaks down. We consider in turn the original formulation and the $T$- and $S$-type axions that arise within an extra-dimensional context.

\subsubsection{Original formulation}

In the initial formulation the UV completion for scales above $f$ was assumed to involve a second scalar that combines with the axion to linearly realize the PQ symmetry as a complex scalar $\Phi$. In this picture the modulus of $\Phi$ acquires a mass proportional to $f \sim \langle \Phi \rangle$ and the axion starts life as the phase of $\Phi \propto e^{i \axion/f}$. 

Motivated by string theory and black-hole thought experiments it is then assumed that UV physics cannot support an unbroken global symmetry, and so at some large scale $M$ the form of the scalar potential for $\Phi$ cannot be assumed to be invariant under re-phasings of $\Phi$. As an expansion in powers of $\Phi$, the generic potential form would be
\be
  V_\UV(\Phi)= \frac{M^4}{2}\sum_{n=1}^{\infty}\left(c_{n}\frac{\Phi^{n}}{M^{n}}+ \hbox{h.c.} \right)\,,
\label{eq:DeltaV}
\ee
where the $c_{n}$'s are in general complex. This is true even if the UV physics is assumed to be CP-invariant because $c_{n}$ will inherit the phase of the fermion mass matrix after chiral PQ rotations. In the initial formulation $M$ is assumed to be the Planck mass $M_p$, and although we can see that such a choice would dominate smaller $M$ for the terms with $n < 4$ it is likely that $M < M_p$ would be more dangerous for $n > 4$. Early workers typically assumed that the renormalizable part of the potential would be tuned to make the axion potential sufficiently shallow and so effectively started the sum in \pref{eq:DeltaV} at $n = 5$.

Freezing the modulus field at $\langle \Phi \rangle = f$ and integrating it out at the classical level leads to the following effective axion potential 
\be 
V_\UV(\axion) = \frac{M^4}{2} \sum_{n=1} |c_{n}|\frac{f^{n}}{M^{n}} \left(e^{i\delta_{n}} \, e^{in {\axion}/{f}}+\hbox{h.c.} \right)  = M^4\sum_{n=5} |c_{n}|\frac{f^{n}}{M^{n}}\cos \left(\frac{n \axion}{f}+\delta_{n} \right)\,,
\label{eq:DeltaV2}
\ee
where we shift fields so that the standard QCD solution is $\axion = 0$. The QCD minimum therefore remains unchanged if $V_\UV'(0) = 0$ and this would be true if all of the $\delta_n$'s were to vanish. Although the axion potential height (and therefore possibly axion mass) might still change due to the presence of $V_\UV(\axion)$, evasion of the strong CP problem requires only that the minimum for $\axion$ remains unmoved.

\bigskip\noindent
{{\it Stability of the minimum}}: 

\medskip\noindent
For $\delta_{n}\,, |c_n| \sim \cO(1)$ we can estimate the size of the effective value of $\bar \theta_{\rm eff}$ by perturbing around the QCD minimum at $\axion = \axion_\QCD$:
\be
\bar{\theta}_{\rm eff}\simeq - \frac{V'_\UV(\axion_{\QCD})}{f V''_\QCD(\axion_{\QCD})} \sim  \frac{V_\UV(\axion_{\QCD})}{ V_\QCD(\axion_{\QCD})} \sim \frac{M^4}{\Lambda_\QCD^4} \left( \frac{f}{M} \right)^{n_0} \,,
\label{eq:newtheta}
\ee
where $n_0$ represents the first power appearing in the sum. For example, requiring $\bar{\theta}_{\rm eff}<10^{-10}$ for the example $f=10^{12}$ GeV, $M = M_p = 10^{18}$ GeV  and $\Lambda_\QCD \simeq 0.2$ GeV  in (\ref{eq:newtheta}) requires $n_0 \gsim 15$.

\bigskip\noindent
{{\it Stability of the axion mass}}: 

\medskip\noindent
The change to the axion mass induced by the UV axion potential is given by
\be
\delta m_a^2 = \left. \frac{\partial^{2}V_\UV(\axion)}{\partial \axion^{2}} \right\vert_{\axion=0}= M^2 \sum_{n=1} n^{2} |c_{n}|\left(\frac{f}{M}\right)^{n-2} \cos \delta_{n} \,,
\label{eq:mass}
\ee
which can be significant unless the coefficients $|c_{n}|$'s are extremely small even if all the $\delta_n$'s could be contrived to vanish. 
When significant such contributions spoil the relation $m_{a}f \sim m_{\pi}F_{\pi}$ that holds for the low-energy QCD contribution and on which most axion phenomenology is based. Because the mass is not inversely proportional to $f$ this expression shows that the relation between $m_{a}$ and $f$ need not be inversely proportional to one other, for example allowing a very heavy axion to be still very weakly coupled to matter -- a drastic change relative to standard axion phenomenology.  

\subsubsection{$T$-type axions}

The story is similar for $T$-type axions, at least below the Kaluza-Klein scale where they are 4D scalars. No quality issue arises above the KK scale because here the relevant fields are higher-dimensional form fields $H_{\MNP}$ and the only symmetries involved are gauge symmetries like $B \to B + \exd \lambda$ \cite{Kallosh:1995hi}. 

Recalling that $T$-type axions, $b$, arise as extra-dimensional reductions of the form $B_{mn}(x,y) = b(x) \, \omega_{mn}(y)$, with $\omega_{mn}$ a harmonic form in the extra dimensions, the origin of the low-energy shift symmetry $b \to b + c$ (for constant $c$) has its origins as the extra-dimensional transformation $B_{mn} \to B_{mn} +c \, \omega_{mn}$. This is a symmetry of $H = \exd B$ because harmonic forms are closed: $\exd \omega = 0$. It is strictly speaking a `large' gauge transformation because harmonic forms are not exact: there does not globally\footnote{The situation resembles a gauge field $A_m(x,y)$ dimensionally reduced on a circle, so $A_m(x,y+L) = A_m(x,y)$. In this case the massless scalar would be $A_m(x,y) = a(x) \omega(y)$ where $\omega(y)$ is independent of $y$, for which the shift symmetry $a \to a + c$ locally corresponds to a gauge transformation $A_m \to A_m + \partial_m \zeta$ if $\partial \zeta/\partial y = c$, but this cannot be done globally because the solution    cannot satisfy $\zeta(y+L)=\zeta(y)$.} exist a $\lambda_m$ such that $\omega = \exd \lambda$.

The quality problem arises because the shift symmetry in the low-energy 4D theory is not a local gauge symmetry and so it in principle need not be respected by UV effects. One consequently cannot completely preclude the generation of a scalar potential, 
\be
 V_\UV(b) \sim M^4\sum_n c_n \left(\frac{b}{M} \right)^n  \,,
\ee
where $c_n$ are dimensionless order-unity coefficients. But its failure to be a local gauge symmetry is a global obstruction rather than a local one and this means that UV effects cannot generate $V_\UV(b)$ until scales are integrated out that `see' the topology that can distinguish $\omega$ from $\exd \lambda$. This implies two sorts of changes to the standard quality-problem argument. First, the scale $M$ where problems first arise cannot be higher than the KK scale $M \sim 1/R_{\KK}$ corresponding to the size of the 2D cycle in the extra dimensions whose presence is associated with the existence of the harmonic form $\omega_{mn}(y)$. Second, the physics at scale $M$ that generates the potential must itself be sensitive to the nontrivial topology, often leading to additional suppressions.

For instance, an example of physics that can generate PQ-violating operators in (\ref{eq:DeltaV}) identified in \cite{Kallosh:1995hi, Holman:1992ah} is wormhole~\cite{Giddings:1987cg}. For these the coefficients $c_{n}$ in (\ref{eq:DeltaV2}) are exponentially suppressed, given by \cite{Kallosh:1995hi}
\be
c_{n}\sim e^{-S}\sim e^{-(M_p L)^{2}}
\label{eq:cn}
\ee
where $S$ is a wormhole action and $L$ the size of its throat. Maintaining the success of the PQ mechanism requires $S\gsim190$. More complicated configurations are possible for extra-dimensional theories, for which $M_p$ can be replaced by another UV gravity scale $M_g$, that might be the string scale or the extra-dimensional Planck scale $M_g$ in specific examples. Similarly $L$ can be one of the geometric scales of the background, that could (but need not) be approximately a compactification scale $R_{\KK}$. All known semiclassical arguments of this type must assume $M_g L \gg 1$ for the calculation to be under control, because semiclassical methods are justified within an expansion in powers of $(M_g L)^{-1}$ within any gravitational EFT. $M_g L \sim 14$ suffices to ensure $S\gsim190$ and so satisfying this constraint seems not that difficult within the semiclassical regime. These kinds of arguments were used  in \cite{NaturalnessForms} to argue for the absence of large gravitational correction to the inflaton potential. 

\subsection{The dual Quality Problem}

For $S$-type axions the representation directly obtained from UV physics is the field  $b_{\mu\nu}$ dual to the scalar axion. And as alluded to earlier -- {\it c.f.} \S\ref{ssec:SPot} -- issues of UV sensitivity can look very different in dual formulations to scalar theories, with for example the existence of a dual implying that the effective couplings for terms like $\axion^n \in V_\UV(\axion)$ come suppressed by powers of the axion mass $(m/M)^{n}$ relative to generic scalar estimates. Such suppressions can be enormous given the small size of $m$ relative to UV scales. 

We therefore revisit earlier discussions of how the axion quality problem arises in the dual formulation, partly motivated by recent discussions \cite{Sakhelashvili:2021eid, Dvali:2022fdv} that argue that gravity causes new problems. Although we confirm the important role played by multiple 3-form potentials \cite{DualStrongCP} in the framing of the dual quality problem, we also show that the many 3-forms found in string vacua do not generically pose a problem. Problems are only caused where strongly interacting systems make instanton-like effects important and this is not the case for the many `elementary' 3-forms that descend from extra dimensional vacua. We argue that for similar reasons 4D gravitational Chern-Simons forms also need not cause problems (such as for string vacua where the UV completion of gravity is described by weakly coupled physics). 

To the extent that the shape of the axion potential $V(\axion)$ is dual to interactions like $W(X)$ involving the 4-form field strength $X = \frac{1}{4!} H_{\mu\nu\lambda\rho} \epsilon^{\mu\nu\lambda\rho}$, one might think that the dual version of the axion quality issue should hinge on the detailed form of UV contributions to $W(X)$. This proves not to be right, as we now argue. The central point turns on the Legendre transformation relating $V(\axion)$ to $W(X)$; in particular on \pref{alphavsWX2} and \pref{dVdalphavsX2}, that state
\be \label{alphavsWX22}
    \left( \frac{\partial W}{\partial X} \right)_{H = \cH} = m\, \axion - \ol\theta \tilde \Lambda_\QCD^2
    \quad \hbox{and} \quad
     \frac{\partial V}{\partial \axion} =  mX\,.
\ee

On the scalar side the strong-CP problem is {\it not} solved unless $m \axion = \ol\theta \tilde \Lambda^2$ at the minimum of $V$, and the quality problem is the statement that corrections to $V$ can perturb the minimum so that this relation fails. Although $X$ always vanishes at a minimum for $V$, eq.~\pref{alphavsWX22} suggests that on the dual side the criterion for satisfying the strong-CP problem is that $\partial W/\partial X = 0$ is satisfied when $X = 0$. So the quality problem seems to hinge on whether or not UV physics can introduce a linear term $\delta W = \eta  X$ whose inclusion would modify \pref{alphavsWX22} in a way that obstructs having $m\, \axion = \ol\theta \tilde \Lambda_\QCD^2$ be a solution to $\partial V/\partial \axion = 0$.

Suppose, then, that one finds after integrating out the UV physics an EFT below the QCD scale of the form \pref{BelowQCD}, but with a linear term in $X$ whose coefficient is {\it not} proportional to the CP violating parameter $\ol\theta$:
\bea \label{BelowQCD2}
   \cL_1(C,B ) &=& - \frac{1}{2\cdot 3!} (G_{\mu\nu\lambda} + mC_{\mu\nu\lambda})( G^{\mu\nu\lambda} + mC_{\mu\nu\lambda})  - \frac{1}{3!} \, \epsilon^{\mu\nu\lambda\rho} ( G_{\mu\nu\lambda} + m C_{\mu\nu\lambda})  J_\rho \nn\\
   && \qquad- \frac{1}{4!} (\ol\theta + \eta) \, \tilde \Lambda_\QCD^2 \epsilon^{\mu\nu\lambda\rho} H_{\mu\nu\lambda\rho}       - \frac{1}{2\cdot 4!} H_{\mu\nu\lambda\rho} H^{\mu\nu\lambda\rho}  + \cdots\,,
\eea
with two low-energy CP-violating parameters $\ol\theta$ and $\eta$. Dualizing this system as above then shows that scalar potential on the scalar side is given by a function of $m \axion - (\ol\theta +\eta) \tilde \Lambda^2$, in which $\ol\theta$ and $\eta$ only appear as a sum. The arguments of \S\ref{ssec:DualQCDPQ} now show that this potential is minimized when $m \axion - (\ol\theta +\eta) \tilde \Lambda^2 = 0$. Repeating the calculation of the neutron electric dipole moment (edm) in this case -- for a recent review, see for example \cite{Hook:2018dlk} -- then shows that the neutron edm also depends only on the sum $\ol\theta + \eta$ and so would continue to vanish when $\axion$ is evaluated at the potential's minimum. Interestingly, just introducing new terms linear in $H_{\mu\nu\lambda\rho}$ in \pref{BelowQCD} appears {\it not} to cause a quality problem. 

\subsubsection{A second strong sector}
\label{sssec:SecondStrong}

Just introducing a linear term in $H_{\mu\nu\lambda\rho}$ in \pref{BelowQCD} does  not  cause a quality problem because doing so below the QCD scale is like introducing the new CP-violating parameter $\eta$ only in the $F\wedge F$ term of \pref{QCDAxialDualNoAlpha} above the QCD scale ({\it i.e.}~shifting $\theta \to \theta + \eta$). This also would not cause a quality problem on the scalar side. For there to be a problem requires there to be a CP-violating contribution to $V(\axion)$ that is {\it independent} of the CP-violation in the $\theta$-term. 

What might this look like on the dual side? One way to proceed is to imagine a specific type of CP-violating UV completion and ask what happens in this case. One such an example would add another strongly interacting nonabelian gauge sector that also contributes to the axion anomaly. In this case $V_\UV(\axion)$ is obtained by integrating out the new gauge sector and this is by construction independent of the QCD-generated part. A dual formulation of this type of system would involve a new Chern-Simons form $E_{\mu\nu\lambda}$ for the new sector in addition to the QCD field $C_{\mu\nu\lambda}$, since both gauge sectors have their own Chern-Simons fields and either of these can be the field that is eaten by $B_{\mu\nu}$. Instead of \pref{BelowQCD2} below the QCD scale one would find the following low-energy action
\bea \label{BelowQCD3}
   \cL_1(C,E, B ) &=& - \frac{1}{2\cdot 3!} G_{\mu\nu\lambda}   G^{\mu\nu\lambda}   - \frac{1}{3!} \, \epsilon^{\mu\nu\lambda\rho}   G_{\mu\nu\lambda}   J_\rho  
    - \frac{1 }{4!}  \epsilon^{\mu\nu\lambda\rho}  \Bigl(\ol\theta \tilde \Lambda_\QCD^2 H_{\mu\nu\lambda\rho}  + \eta \tilde \Lambda_\ssX^2   K_{\mu\nu\lambda\rho} \Bigr)  \nn\\
   && \qquad\qquad - \frac{1}{2\cdot 4!} \Bigl( H_{\mu\nu\lambda\rho} H^{\mu\nu\lambda\rho} + K_{\mu\nu\lambda\rho} K^{\mu\nu\lambda\rho} \Bigr) + \cdots\,, 
\eea
where $K = \exd E$ and $H = \exd C$ and $G = \exd B + m C + \tilde m E$. 

Proceeding as before we introduce a Lagrange multiplier $\axion$ to enforce the $G$ Bianchi identity and then semiclassically integrate out $G$, $H$ and $K$ to find
\be
   \cL_2(  \axion)= - \frac12 (\partial \axion)^2    - J^\mu \partial_\mu \axion - \frac12 J_\mu J^\mu  - V(\axion)  \,,
\ee
where defining $X = \frac{1}{4!} \epsilon^{\mu\nu\lambda\rho}H_{\mu\nu\lambda\rho}$ and $Y = \frac{1}{4!} \epsilon^{\mu\nu\lambda\rho}K_{\mu\nu\lambda\rho}$ we find
\be\label{V2Forms}
  V(\axion) = - W(X,Y) + (m \axion - \ol\theta \tilde \Lambda_\QCD^2) X + (\tilde m \axion - \eta \tilde \Lambda_\ssX^2) Y  \,,
\ee
where $W = \frac12(X^2 + Y^2) + $(higher powers). At the saddle point $(H,K) = (\cH,\cK)$ we have
\be \label{alphavsWX26}
    \left( \frac{\partial W}{\partial X} \right)_\ssY = m\, \axion - \ol\theta \tilde \Lambda_\QCD^2 
    \quad\hbox{and} \quad
    \left( \frac{\partial W}{\partial Y} \right)_\ssX  = \tilde m\, \axion - \eta \tilde \Lambda_\ssX^2 \,,
\ee
where the subscripts indicate what is held fixed in the derivative. Differentiating \pref{V2Forms} implies
\be \label{dVdalphavsX25}
  \frac{\partial V}{\partial \axion} = m X + \tilde m Y\,.
\ee

This does have a quality problem because the competition between the two gauge sectors drives the axion away from the minimum for which the neutron electric dipole moment vanishes. For the simplest example -- where $W = \frac12(X^2 + Y^2)$ -- we can see explicitly how the shift of the global minimum of the axion potential is induced. From \pref{dVdalphavsX25} we learn that $\partial V(\axion)/\partial\axion=0$ takes place at $Y=-(m/\tilde{m})X$. From \pref{alphavsWX26}, we obtain
\be
X=m\axion-\bar{\theta}\tilde{\Lambda}_{QCD}^{2}\quad\hbox{and} \quad Y = - \left(\frac{m}{\tilde{m}}\right) X = \tilde{m}\axion-\eta\tilde{\Lambda}_{X}^{2} 
\label{eq:X}
\ee
Equating these two expressions for $X$ and solving for $\axion$, we obtain
\be
\axion_{\rm min} = \frac{m\bar{\theta}\tilde{\Lambda}_{\QCD}^{2}+\tilde{m}\eta\tilde{\Lambda}_{\ssX}^{2}}{m^{2}+\tilde{m}^{2}} = \frac{\axion_{\QCD}+ ({\tilde{m}\eta\tilde{\Lambda}_{X}^{2}}/{m^{2}})}{1+\left({\tilde{m}}/{m}\right)^{2}}\simeq \axion_{\QCD}+\frac{\tilde{m}\eta\tilde{\Lambda}_{X}^{2}}{m^{2}}\,,
\ee
which denotes the global minimum before introducing an extra three form gauge field by $\axion_{\QCD}=\bar{\theta}\tilde{\Lambda}_{QCD}^{2}/m$. The approximate equality assumes $m \gg \tilde{m}$ so as not to spoil the QCD axion solution the strong CP problem. 

Finally, defining the UV contribution to the effective vacuum angle by $\ol\theta_{\rm eff} := (\axion_{\rm min} - \axion_\QCD)/f$ where $m f \simeq \tilde{\Lambda}_{QCD}^{2}$, we obtain the constraint
\be \label{eq:Deltathetamin}
 \bar{\theta}_{\rm eff} \sim \eta\left(\frac{\tilde{m}}{m}\right)\left(\frac{\tilde{\Lambda}_{X}}{\tilde{\Lambda}_{QCD}}\right)^{2} \lsim 10^{-10} \,.
\ee
Although this derivation assumed the simplest form $W = \frac12(X^2 + Y^2)$, the reasoning presented here can be applied to a more complicated $W(X,Y)$. In such a case \pref{dVdalphavsX25} remains unchanged while \pref{alphavsWX26} and \pref{eq:X} are modified. But $\axion_{\rm min}$ remains connected to the value for $(X,Y)$ that makes $\partial V/\partial\axion$ vanish via \pref{alphavsWX26} and \pref{dVdalphavsX25}. Once $\axion_{\rm min}$ is expressed in terms of $\axion_{\QCD}$, one can always infer $\bar{\theta}_{\rm eff}$ as above and impose the constraint $\bar{\theta}_{\rm eff}<10^{-10}$.

The upshot is this: the requirement of multiple strongly coupled sectors on the dual side to generate a quality problem is much more explicit because the contribution of each sector is described by a separate 3-form potential, rather than having everything all be rolled into the same scalar potential. 

\subsubsection{Multiple fundamental 3-forms}

At first sight the previous section makes it sound like string theory should typically have a huge quality problem, because of the generic  appearance there of multiple 3-form potentials. We identify the circumstances under which these potentials could cause a quality problem and argue why such a problem generically does not happen. We also discuss how these criteria bear on a recent realization of these issues \cite{Sakhelashvili:2021eid, Dvali:2022fdv}.

To start consider how the EFT \pref{BelowQCD3} above the QCD scale would be modified by the presence of many 3-form potentials $\cC^\ssA_{\mu\nu\lambda}$ (where $A = 1, \dots , N$ distinguishes the different UV potentials):
\bea \label{QCDAxialDualNoAlphaz}
   \cL_1(B,A, \cC) &=&  - \frac{1}{2\cdot 3!} G_{\mu\nu\lambda} G^{\mu\nu\lambda}    - \frac{1}{3!} \, \epsilon^{\mu\nu\lambda\rho}  G_{\mu\nu\lambda}  J_\rho    - \frac14 F_{\mu\nu} F^{\mu\nu}- \frac{\theta}{2} \, \epsilon^{\mu\nu\lambda\rho} F_{\mu\nu}F_{\lambda\rho}  \nn\\
   && \qquad\qquad  - \frac{1 }{4!} \, \eta_\ssA \cH^\ssA_{\mu\nu\lambda\rho}   \epsilon^{\mu\nu\lambda\rho}    - \frac{1}{2\cdot 4!}   \cH^\ssA_{\mu\nu\lambda\rho} \cH_\ssA^{\mu\nu\lambda\rho}  + \cdots  \,.
\eea
where $\cH^\ssA = \exd \cC^\ssA$ and $G = \exd B + S$ for the QCD Chern-Simons 3-form that satisfies $\Omega = \exd S$ with $\Omega$ as given in \pref{OmegaDefFF}. To the extent that none of the new fields $\cH^\ssA_{\mu\nu\lambda\rho}$ appear in the Bianchi identity $\exd G = \Omega$ they do not couple to QCD or to $B_{\mu\nu}$ and so play no role in the duality transformation from $B_{\mu\nu}$ to $\axion$. One then arrives below the QCD scale with the lagrangian
\bea \label{BelowQCDz}
   \cL_1(\cC,B ) &=& - \frac{1}{2\cdot 3!} G_{\mu\nu\lambda}  G^{\mu\nu\lambda}  - \frac{1}{3!} \, \epsilon^{\mu\nu\lambda\rho}  G_{\mu\nu\lambda}  J_\rho - \frac{\ol\theta}{4!} \, \tilde \Lambda_\QCD^2 \epsilon^{\mu\nu\lambda\rho} H_{\mu\nu\lambda\rho}       - \frac{1}{2\cdot 4!} H_{\mu\nu\lambda\rho} H^{\mu\nu\lambda\rho}  \nn\\
   && \qquad\qquad  - \frac{1 }{4!} \, \eta_\ssA \cH^\ssA_{\mu\nu\lambda\rho}   \epsilon^{\mu\nu\lambda\rho}    - \frac{1}{2\cdot 4!}   \cH^\ssA_{\mu\nu\lambda\rho} \cH_\ssA^{\mu\nu\lambda\rho}  + \cdots\,.
\eea

Dualization proceeds as before, with the introduction of the scalar $\axion$ to enforce $\exd G = \Omega$, and the saddle point in the integral over the 3-form potentials becomes
\be \label{alphavsWXz}
    \left( \frac{\partial W}{\partial X} \right)_{\ssY} = m\, \axion  - \ol\theta \tilde \Lambda_\QCD^2 \quad \hbox{and} \quad
    \left( \frac{\partial W}{\partial Y^\ssA} \right)_\ssX =- \eta_\ssA \,,
\ee
where 
\be \label{Wform2}
   W = \frac12 X^2 + \frac12 Y^\ssA Y_\ssA  + \hbox{(higher powers)} \,,
\ee
and we define as before $X = \frac{1}{4!} \epsilon^{\mu\nu\lambda\rho}H_{\mu\nu\lambda\rho}$ and $Y^\ssA = \frac{1}{4!} \epsilon^{\mu\nu\lambda\rho} \cH^\ssA_{\mu\nu\lambda\rho}$. The dual lagrangian is
\be
   \cL_2(  \axion)= - \frac12 (\partial \axion)^2    - J^\mu \partial_\mu \axion - \frac12 J_\mu J^\mu  - V(\axion)  \,,
\ee
where 
\bea\label{V2Formsz}
  V(\axion) &=& - W(X,Y^\ssA) + (m \axion - \ol\theta \tilde \Lambda_\QCD^2) X  - \eta_\ssA Y^\ssA  \nn\\
  &=& -\frac12 \, X^2 + (m \axion - \ol\theta \tilde \Lambda_\QCD^2) X  + \frac12 \eta_\ssA \eta^\ssA  \,,
\eea
and so
\be \label{dVdalphavsX25z}
  \frac{\partial V}{\partial \axion} = m X  \,.
\ee
We see that $X = 0$ in the vacuum and this implies from \pref{alphavsWXz} and \pref{Wform2} that the strong-CP problem remains solved. 

These arguments also show that two ingredients are required for additional 3-form potentials to cause a problem:
\begin{enumerate}
\item The additional 3-form potential $\cC^{\ssA_0}$ must contribute to the Bianchi identity for $G$, and so $\kappa_{\ssA_0} \neq 0$ in the expression $\exd G =\Omega + \kappa_\ssA \cH^\ssA$, where $\cH^\ssA = \exd \cC^\ssA$; and 
\item The additional 3-form potential must appear linearly in $W$, so $\eta_{\ssA_0} \neq 0$ in \pref{BelowQCDz}.
\end{enumerate} 
When both of these are satisfied then $\axion$ couples to $\cH^\ssA$ and leads to the competition of minima as in \pref{alphavsWX26} along the lines described in \S\ref{sssec:SecondStrong}. The need for both of these conditions to be true is why the bound \pref{eq:Deltathetamin} is proportional to both $\eta$ and $\tilde \Lambda_\ssX^2$. The good news is that the vanishing of $\kappa_\ssA$ can be enforced by a gauge symmetry, since $\kappa_\ssA$ can only be nonzero if $B$ transforms as $B \to B - \kappa_\ssA \Lambda^\ssA$ under the 3-form gauge transformations $C^\ssA \to C^\ssA + \exd \Lambda^\ssA$. 

There is at least one example of a 3-form potential which we know must exist and which also contributes to the Bianchi identity $\exd G$: the gravitational Chern Simons 3-form, $S_g$.   The existence of a PQ-Lorentz-Lorentz anomaly requires this form to appear in $G$ and so have a nonzero coefficient $\kappa_g$ in the same way that the PQ-QCD-QCD anomaly requires the QCD Chern Simons form to appear there. Ref.~\cite{DualStrongCP} argues that this is real trouble whose evasion requires model-building, such as that done in \cite{Dvali:2022fdv}. 

Whether the existence of this form is a problem or not depends on whether it also satisfies item 2 above: {\it i.e.}~whether or not it appears linearly in the lagrangian with coefficient $\eta_g \neq 0$. How big should $\eta_g$ be expected to be? Because any 4-form field strength $\cH = \exd \cC$ is locally a total derivative it wants to drop out of perturbative physics when it appears linearly in the action (much as does $F\wedge F$). Consequently its appearance in a low-energy action requires some sort of nonperturbative process (like an instanton) to contribute to physical processes. This is indeed what happens for QCD for which the linear term in $\Omega$ appears with coefficient   
\be
 \tilde \Lambda_\QCD^2 \propto M^2 \, e^{-  2\pi b/\alpha}
\ee
with $M$ a UV scale, $b$ a pure number and $\alpha = g^2/4\pi$ the QCD coupling. The tell-tale nonperturbative dependence on $\alpha$ is a semiclassical consequence of the topological character of $\int F\wedge F$ and $\int H$. 

This suggests that for gravity a linear term in $\cH^g$ should similarly be of size
\be
  \eta \propto  M^2 \, e^{- (M L)^2} 
 \label{eq:eta} 
\ee
for a characteristic instanton length scale $L$ and gravitational UV scale $M$ given that $(M L)^{-2}$ plays the role of the semiclassical expansion parameter (compare to \pref{eq:cn}). This can be extremely small within the domain of validity of semiclassical reasoning, for which $M L \gg 1$ (as would presumably apply when the UV completion is weakly coupled, such as for perturbative string vacua). 

Examples of three forms characterized by $\eta$ in \pref{eq:eta} include Eguchi-Hanson instantons
\cite{Gibbons:1978tef, Eguchi:1978gw} and the gravitational Chern-Simons 3-form made up of gravitational connection. For the Chern-Simons 3-form ref.~\cite{DualStrongCP} argues that gravity indeed becomes strong in the UV, as would be required for $\eta$ to be significant. This could well be true, but the evidence for there being a problem hinges on how convinced one is about gravitational interactions becoming strong in the UV.

\subsubsection{Multiple-axion solution}

We close this section by remarking that having multiple axion candidates (as is often true for string vacua) can alleviate the above problem associated with multiple 3-form fields, even if the above two conditions are satisfied.\footnote{The use of multiple axions to solve the quality problem is mentioned also in \cite{Heidenreich:2020}, who have different but related motivations for there being a plethora of form fields present in the UV.} This observation points to an equally general quality control mechanism on the scalar side of the duality as well.

To see why, we introduce a second Kalb-Ramond field $\mfB_{\mu\nu}$ to the model of \S\ref{sssec:SecondStrong}, and supplementing the lagrangian of \pref{BelowQCD3} with the appropriate additional kinetic term gives 
\bea \label{BelowQCD3q}
   \cL_1(C,E, B ,\cB) &=& - \frac{1}{2\cdot 3!} \mfG_{\mu\nu\lambda}   \mfG^{\mu\nu\lambda}  - \frac{1}{2\cdot 3!} G_{\mu\nu\lambda}   G^{\mu\nu\lambda}   - \frac{1}{3!} \, \epsilon^{\mu\nu\lambda\rho}   G_{\mu\nu\lambda}   J_\rho \\
   && \qquad - \frac{1 }{4!}  \epsilon^{\mu\nu\lambda\rho}  \Bigl(\ol\theta \tilde \Lambda_\QCD^2 H_{\mu\nu\lambda\rho}  + \eta \tilde \Lambda_\ssX^2   K_{\mu\nu\lambda\rho} \Bigr)  - \frac{1}{2\cdot 4!} \Bigl( H_{\mu\nu\lambda\rho} H^{\mu\nu\lambda\rho} + K_{\mu\nu\lambda\rho} K^{\mu\nu\lambda\rho} \Bigr) + \cdots\,,    \nn
\eea
where as before $K = \exd E$ and $H = \exd C$ and $G = \exd B + m C + \tilde m E$, but now also
\be
   \mfG := \exd \mfB + m_\star E \,.
\ee

This system dualizes much as before: we introduce Lagrange multipliers $\axion$ and $\baxion$ to enforce the $G$ and $\mfG$ Bianchi identities $\exd G = m H + \tilde m K$ and $\exd \mfG = m_\star K$ and then integrate out $G$, $\mfG$, $H$ and $K$ to find
\be
   \cL_2(  \axion)= - \frac12 (\partial \baxion)^2   - \frac12 (\partial \axion + J)^2     - V(\axion, \baxion)  \,,
\ee
with
\be\label{V2Forms2}
  V(\axion,\baxion) = - W(X,Y) + (m \axion - \ol\theta \tilde \Lambda_\QCD^2) X + (\tilde m \axion + m_{\star}\baxion - \eta \tilde \Lambda_\ssX^2)Y \,,
\ee
and we define as before $X = \frac{1}{4!} \epsilon^{\mu\nu\lambda\rho}H_{\mu\nu\lambda\rho}$ and $Y = \frac{1}{4!} \epsilon^{\mu\nu\lambda\rho}K_{\mu\nu\lambda\rho}$. For the simplest example of $W = \frac12(X^2 + Y^2)$, at the saddle point $(H,K) = (\cH,\cK)$ gives the following relation between $(X,Y)$ and $(\axion,\baxion)$:
\be \label{alphavsWX262}
    \left( \frac{\partial W}{\partial X} \right)_\ssY = m\, \axion - \ol\theta \tilde \Lambda_\QCD^2 
    \quad\hbox{and} \quad
    \left( \frac{\partial W}{\partial Y} \right)_\ssX  = \tilde m\, \axion +m_{\star}\, \baxion - \eta \tilde \Lambda_\ssX^2 \,.
\ee

Differentiating \pref{V2Forms2} with respect to $\axion$ and $\baxion$ implies
\be \label{dVdalphavsX252}
  \frac{\partial V}{\partial \axion} = m X + \tilde m Y,\qquad \frac{\partial V}{\partial \baxion} = m_{\star} Y\,.
\ee
and so shows that all extrema of the potential satisfy $X =  Y = 0$ (provided $m, \tilde m$ and $m_\star$ are nonzero).
Because $\partial W/\partial X$ vanishes at $X = 0$ it follows that the dynamics chooses $\axion_{\rm min}$ to satisfy $\ol{\theta}\tilde{\Lambda}_{\QCD}^{2}/m = \ol{\theta}f$ through \pref{alphavsWX262}; the axion quality problem essentially disappears.

What happened? Why does introducing another axion resolve the quality problem? The crux of the mechanism lies in the difference between eq.~\pref{dVdalphavsX252} and \pref{dVdalphavsX25}. The derivative of the potential always sets a linear combination of 4-form field strengths to zero and if there are as many equations as there are fields the only solution is generically to have all 4-form field strengths vanish. Once this is true then the first of eqs.~\pref{alphavsWX262} ensures that this solution solves the strong-CP problem. Trouble only arises -- as it did in \S\ref{sssec:SecondStrong} -- when there are fewer equations than unknowns ({\it i.e.}~fewer axions than 3-form potentials), since then $X$ need not vanish and eqs.~\pref{alphavsWX262} become competing conditions on the same axion variable.

A similar mechanism also exists on the scalar side of the duality. If two sectors generate contributions to the QCD axion potential then the problem arises because these compete in the value they imply for the axion expectation value. Introducing a second anomalous $U(1)$ symmetry that also has anomalies with the same two sectors provides enough latitude to minimize each sector's potential separately, thereby removing the troublesome competition. 

For instance, suppose there was a new non-Abelian gauge sector $\mathcal{G}$ and suppose the usual PQ symmetry has both a QCD anomaly and an anomaly in the $\cG$ sector. This is the kind of thing that can cause a quality problem because of the contradictory conditions the two sectors impose on the QCD axion. But also introducing another global $U(1)$ with only a $\cG$-sector anomaly can help because there is a linear combination of the PQ symmetry and the new $U(1)$ that is anomaly free in the $\cG$ sector and the PQ mechanism then goes through using this new symmetry.\footnote{Ref.~\cite{Dvali:2022fdv} uses a special case of this general mechanism by introducing an extra $U(1)$ symmetry in the leptonic sector to resolve the problem raised by the assumption that gravity is strongly coupled.}

\section{UV completion and matter couplings}
\label{sec:SemiClass}

Since the motivations both for considering Kalb-Ramond fields and for the absence of global symmetries come from the UV, it is useful to ask whether there are other potential surprises for axion physics having their roots in the UV. This section examines two such examples; one each for $T$-type and for $S$-type axions. For $T$-type axions we provide simple examples for which physical axion-matter couplings like $g_{\aff}$ can be much smaller than the naive value $1/f_b$ read off from the axion kinetic term. In the example shown here gauge invariant matter couplings like $g_{\aff}$ are order $1/M_p$ despite $f_b$ being an ordinary particle physics scale, while anomalous gauge couplings remain order $1/f_b$ in size (if they exist at all).

For $S$-type axions we show that the corresponding physical couplings indeed are of order $1/f_a$ and we identify the UV physics to which couplings of size $E/f_a$ match at energies $E \gsim f_a$. We also show how $S$-type axions can be examples of weak/strong duality, and that it is the Kalb-Ramond side of the duality that is usually weakly coupled. Weak/strong coupling interchange due to duality could be relevant to applications for which the effects of scalar axions are explored using semiclassical reasoning, and if so would provide a further motivation for taking the Kalb-Ramond formulation as primary. 

\subsection{Extra-dimensional UV completion}

To this end suppose that both Kalb-Ramond field and the standard model arise within an extra-dimensional model. For concreteness' sake we take the higher-dimensional kinetic term for the 2-form field and the Einstein-Hilbert part of the action to be\footnote{For simplicity we ignore extra-dimensional warping in this discussion. We also do not canonically normalize $B_{\MN}$, which here is taken to be dimensionless.}
\be \label{APPBMNkin}
   S_{\rm kin} = -\frac{1}{2}\, M^{2+d} \int \exd^4 x \,\exd^d y \; \sqrt{- \tilde g_{(\ssD)}} \left(  \widetilde\cR + \frac{1}{3!} \, e^{-\lambda \phi} \, G_{\MNP} G^{\MNP}  \right)\,,
\ee
where there are $D=4+d$ spacetime dimensions and $M$ is a UV scale -- the higher-dimensional Planck scale. $\widetilde\cR$ here denotes the Ricci scalar and $\tilde g_{(\ssD)}$ is the determinant of the full $D$-dimensional metric $\tilde g_{\MN}$. As above $H = \exd B + \cdots$ is the Kalb-Ramond field strength and $\phi$ is the extra-dimensional dilaton that often arises within the higher-dimensional gravity supermultiplet. The parameter $\lambda$ depends on higher-dimensional details, with $(d,\lambda) = (2,2)$ for chiral 6D supergravity \cite{6Dsugra}, $(d,\lambda) = (6,1)$ for Neveu-Schwarz 2-forms in 10D supergravity and $(d,\lambda = -1)$ for Ramond 2-forms in 10D supergravity \cite{10Dsugra} (for example).

The derivation of this type of lagrangian as the low-energy limit of a string vacuum usually relies on two approximations: the low-energy approximation (or $\alpha'$ expansion) where energies are well below the string scale $E \ll M_s$; and the weak string coupling approximation, which involves expanding in powers of $e^\phi \ll 1$. For simplicity we restrict ourselves to this limit as well, and specialize to the simplest case $(d,\lambda) = (2,2)$ corresponding to 6D chiral supergravity.

Dimensional reduction to 4D proceeds by integrating out the two extra dimensions and putting the 4D Einstein-Hilbert term into standard form (4D Einstein frame) by appropriately rescaling the 4D part of the metric
\be
   \tilde g_{\mu\nu} = \left( \frac{\cV_{2\star}}{\cV_2} \right) g_{\mu\nu} = \frac{ 1 }{  \cV_2} \left( \frac{M_p^2}{M^2} \right) g_{\mu\nu}  \quad\hbox{where}\quad \cV_d :=  M^d \int \exd^d y \sqrt{\tilde g_{(d)} }
\ee 
is the dimensionless extra-dimensional volume and the subscript `$\star$' on a field denotes its present-day value\footnote{The factor of $\cV_{2\star}$ ensures the rescaling is trivial at present, as required to not change present-day units of length.} and the 4D Planck massis is defined by 
\be
    M_p^2 =  \cV_{2\star} \, M^2  \,.
\ee

\subsubsection*{$S$-type axion}

The kinetic term for $b_{\mu\nu}$ in 4D Einstein frame that is obtained by dimensional reduction is
\be \label{APPhmunurhokin}
  \cL_{\rm kin} =  -\frac1{12}\, M^{2} \cV_2    \sqrt{- \tilde g_{(4)}}   \; e^{-2 \phi} \tilde g^{\mu\nu}  \tilde g^{\beta\rho}  \tilde g^{\xi\zeta} \, H_{\mu\beta \xi} H_{\nu\rho\zeta} =  -\frac{M^4}{12 M_p^2} \sqrt{-g} \; e^{-2 \phi}\,\cV_2^2 \, h_{\mu\nu\beta} h^{\mu\nu\beta} \,,
\ee
where $h_{\mu\nu\lambda} = \partial_\mu b_{\nu\lambda}$ + (cyclic). This last form can be written in terms of a scalar by dualizing as in earlier sections, imposing the Bianchi identity\footnote{The factors of $M$ here are chosen so that $\Omega$ has dimension (mass)${}^4$.} $\exd h = \Omega/M^2$, leading to the dual result
\be \label{STypeKin}
  \cL_{\rm dual} =  -  \sqrt{-g} \left[  \frac{M_p^2 e^{2 \phi}}{\cV_2^2} \, \partial_\mu \mfa \, \partial^\mu \mfa  +  \frac{1}{3!}  \, \mfa \, \epsilon^{\mu\nu\beta\rho}  \Omega_{\mu\nu\beta\rho} \right]  
\ee
which suggests its decay constant can be written $f_a = (M_p/\cV_{d\star}) \, e^{ \phi_\star }$. 

Two things are noteworthy here. First, notice that the volume dependence means that $f_a$ can be very much smaller than Planckian size. In the extreme case of two large extra dimensions (and working in the weak-coupling regime for which $e^\phi$ is moderately small) the size of the extra dimensions can be as large as $MR_\KK \lsim 10^{14}$ and so $\cV_2 \sim (MR_\KK)^2 \lsim 10^{28}$ can be enormous (potentially allowing $f_a \lll M_p$ to be as small as eV energies).  

Second, notice that although \pref{APPhmunurhokin} has large coefficients when $e^{2 \phi} \ll 1$, the same is not true of the kinetic term in \pref{STypeKin}. This reflects how Kalb-Ramond/axion duality is a weak-strong coupling duality from the point of view of the string coupling $g_s \sim e^\phi$. To the extent that semiclassical expansions rely on the leading action being proportional to the inverse of a small coupling\footnote{When this is true then powers of $g_s^2$ and powers of $\hbar$ are equivalent when evaluating a path integral over $e^{i S_0}$.} -- $\cL_0 = L_0/g_s^2$ -- semiclassical methods should fail for the scalar representation but hold for its dual.

\subsubsection*{$T$-type axion}

For $T$-type axions we use $B_{mn}(x,y) = \mfb(x) \, \omega_{mn}(y)$ where in six dimensions the harmonic form can be taken to be proportional to the extra-dimensional volume form $\varepsilon_{mn}(y)$. Typically $\omega_{mn}$ satisfies a quantization condition that states the integral of $\omega_{mn}$ over the two extra dimensions $\oint_\ssC \omega$ is a pure number, proportional to an integer. Because this result is volume independent it follows that $\omega_{mn} = \cV_2^{-1} \, \varepsilon_{mn}$.  

The kinetic term for the $T$-type scalar $\mfb$ obtained in this way is therefore proportional to
\be\label{bmnkin}
  \cL_{\rm kin}  =     -\frac1{2}\, M^{2} \cV_2    \sqrt{- \tilde g_{(4)}}   \; e^{-2 \phi} \tilde g^{\mu\nu} \partial_{\mu}\mfb \, \partial_{\nu}\mfb \, \cV_2^{-2} = - \sqrt{-g}\; M_p^2 e^{-2\phi} \cV_2^{-2} g^{\mu\nu} \partial_\mu \mfb \,\partial_\nu \mfb  \,.
\ee
Notice that the kinetic term, both here and in \pref{STypeKin}, takes the form
\be \label{SL2Rkin}
  \cL_{\rm kin} =   - \frac12 \sqrt{-g}\,M_p^2 \; \left[ \frac{(\partial \mfb)^2 }{\tau^2}+ \frac{(\partial \mfa)^2 }{\sigma^2} \right]
\ee 
with $\tau = \cV_2 \, e^\phi$ in \pref{bmnkin}, and $\sigma =\cV_2 \, e^{-\phi}$ in \pref{STypeKin}.

\subsection{Coupling strengths}
\label{APPsssec:HiDTtype}

What matters for phenomenology is the couplings of the fields $\mfb$ and $\mfa$ to matter. This is controlled by the size of $\cF$ for axion couplings of the form
\be\label{APPFdef}
  \cL_{\rm ax} =  - \frac{1}{2} \, \partial_\mu \axion \, \partial^\mu \axion -  \frac{1}{\cF} \, \partial_\mu \axion J^\mu   \,,
\ee
where $\axion$ is the canonically normalized axion field and $J^\mu$ is a matter current. $\cF^{-1} = g_{\aff}$ is the axion-fermion current if $J^\mu$ is built from fermion bilinears and $\cF^{-1} \simeq g_{agg}$ or $g_{a\gamma\gamma}$ if $J^\mu$ is the Hodge dual of the QCD or QED Chern-Simons 3-form.

For concreteness' sake we evaluate the size of this coupling in the perturbative semiclassical regime where $\cV_2$ is large and the UV physics is weakly coupled (and so $e^\phi$ small). In this limit we have $f_b \gsim M_p/\cV_2 \gsim f_a$ and both are much smaller than $M_p$. In both cases we will see that $\cF$ can (but need not) be simply given by the corresponding decay constant $f_a$ or $f_b$. 

In higher dimensional constructions very often ordinary matter is localized on a space-filling brane, $\Sigma$, within the extra dimensions. $\Sigma$ could be a four-dimensional 3-brane or a higher-dimensional $p$-brane with $3 \leq p \leq 3+d$. If $p > 3$ then the extra-dimensional part of the brane typically wraps some topological cycle within the extra dimensions, and if this were a two-cycle ({\it e.g.}~if $p = 5$) it would also have an associated harmonic 2-form $\omega_{mn}(y)$ required to ensure that $T$-type axions appear in the low-energy 4D theory. We here explore the simplest case $p=3$.

\subsubsection*{$S$-type axion}

A generally covariant low-dimension interaction between $H_{\MNP}$ and matter fields living on the brane, that is linear in $B_{\MN}$ is\footnote{The first equality shows that this interaction is independent of the metric, and this can also be seen after the second equality from the observation that $\epsilon_{\mu\nu\lambda\rho} = \pm \sqrt{-g}$ and so $\epsilon^{\mu\nu\lambda\rho} = \pm (-g)^{-1/2}$.}   
\be \label{APPawedgeint}
   S_{\rm int} = - \hat c \int_{\Sigma} e^{-\beta \phi} H \wedge J = - \frac{c}{3!} \int \exd^4x \sqrt{-g} \;e^{-\beta\phi}\, \epsilon^{\mu\nu\lambda\rho} h_{\mu\nu\lambda} J_\rho \,,
\ee
where $J_\rho$ is a current built from brane-localized matter fields and $\beta$ is a parameter -- like $\lambda$ in \pref{APPBMNkin} -- that is predicted by any specific extra-dimensional UV completion. The matter current $J_\rho$ has dimension (mass)${}^3$ -- making the coupling parameters $\hat c$ and $c$ dimensionless -- and so could be a fermion bilinear or the Hodge dual of a gauge boson Chern-Simons term (though for the Chern Simons term gauge invariance would require $\beta = 0$). Because this term is covariant without use of the metric it does not acquire factors of $\cV_2$ or $M_p/M$ when going to 4D Einstein frame. 

The dual effective theory for $\mfa$ is then found by adding \pref{APPawedgeint} to the kinetic term \pref{APPhmunurhokin}, imposing the Bianchi identity $\exd G = \Omega/M^2$ and integrating out $h_{\mu\nu\lambda}$, modifying \pref{STypeKin} to
\be \label{STypeKinxxx}
  \cL_{\rm dual} =  -  \sqrt{-g} \left[  \frac{M_p^2 e^{2\phi}}{2\cV_2^2} \, D_\mu \mfa \, D^\mu \mfa  +  \frac{1}{3!}  \, \mfa \, \epsilon^{\mu\nu\beta\rho}  \Omega_{\mu\nu\beta\rho} \right]  
\ee
where
\be\label{APPDmumfadef}
   D_\mu \mfa :=  \partial_\mu \mfa +  \frac{c}{M_p^2}  \, e^{-(\beta + 2)\phi}\, \cV_2^{2} \,  J_\mu  =  \partial_\mu \mfa +  \frac{c}{f_a^2}  \, e^{-\beta \phi} \,  J_\mu   \,.
\ee
As before we use the kinetic term to identify $f_a = (M_p/\cV_{2\star}) \, e^{\phi_\star }$. Using $M_p^2 = M^2 \cV_{2\star}$ with $\cV_{2\star} = (MR_\KK)^2$ for a Kaluza-Klein length scale $R_\KK$, this implies $f_a \sim M\cV_{2\star}^{-1/2} e^{\phi_\star} \sim (1/R_\KK) \, e^{\phi_\star}$, and so $f_a \sim m_\KK \sim 1/R_\KK$ when $e^{\phi_\star}$ is not that much smaller than order unity. 

The physical coupling that comes from comparing the kinetic and $\partial_\mu \mfa \, J^\mu$ term to \pref{APPFdef} is
\be\label{APPFsdef}
  g_{agg} \sim g_{\aff} = \frac{1}{\cF_{\aff}} \sim \frac{c\cV_{2\star} e^{-(\beta+1) \phi_\star}}{M_p} \sim \frac{c\, e^{-\beta \phi_\star}}{f_a}  \quad \hbox{for couplings to $J$} \,.
\ee
In the special case where $J_\mu$ is the Hodge dual of a gauge-field Chern Simons term, gauge invariance also requires we take $\beta = 0$, and once this is done the coupling in \pref{APPFsdef} agrees (up to numerical factors) with the physical coupling to $\Omega$ implied by \pref{STypeKinxxx}. This coupling becomes strong when $E \sim f_a$, which we've seen is of order the Kaluza-Klein scale in the special case of two extra dimensions. 

\subsubsection*{$T$-type axion}

The lowest-dimension generally covariant and gauge invariant interaction that couples $H_{\MNP}$ to matter localized on a space-filling 3-brane and that is linear in the components $H_{\mu mn}$ has the form
\be \label{APPbwedgeint}
   S_{\rm int} = \int_{\Sigma} \; e^{-2\phi}\; {}^\star H  \wedge J  \ni \frac{M_p^2}{M^2}\int \exd^4x \sqrt{-g} \; \frac{e^{-2\phi}}{\cV_2^2} g^{\mu\nu} \partial_\mu \mfb(x) J_{\nu}(x)   \,,
\ee
where ${}^\star H$ denotes the 6D Hodge dual and we choose the $\phi$ coupling to be the same as the kinetic term. 

The kinetic and interactions terms combine to give the effective action (in 4D Einstein frame)
\be
   S_{\rm eff} = \int \exd^4x \, \sqrt{-g} \;  \frac{M_p^2}{\tau^2}\left[ (\partial \mfb)^2 + \frac{\partial_\mu\mfb J^\mu}{M^2} \right]
\ee
where $\tau := \cV_2 \, e^\phi$ as before. Inspection of the kinetic term identifies the decay constant as $f_b \simeq M_p/\tau_\star$ with $\tau_{\star} \propto \cV_{2\star}$ denoting the present value of $\tau$. Because $M_p^2 \simeq M^2 \cV_{2\star}$ we see that $f_b \sim M \cV^{-1/2}_{2\star} \sim R_\KK^{-1}$ is of order the KK scale in size.

Canonically normalizing by rescaling $b = M_p \, \mfb/\tau_{\star}$ ---- then produces a lagrangian of the form \pref{APPFdef} with 
\be
   \cF \sim \frac{M^2 \tau_{\star}}{M_p}
   \sim M_p  \gg f_b \sim \frac{M_p}{\tau_\star}  \,.
\ee
As is typical for KK modes the field $\mfb \in B_{mn}$ couples with gravitational strength. Notice that the ratio $\cF / f_b \propto \cV_{2\star}$ can be enormous, since $\cV_{2\star}$ can be as large as $10^{28}$ in the extreme case of two large eV-scale dimensions. In this case gauge invariance precludes choosing $J$ to be the dual of the Chern Simons form of a brane localized gauge sector, even in the absence of any $\phi$-dependence in \pref{APPbwedgeint}. 
 
It is not that surprising to have a breakdown of 4D EFT methods at the KK scale, but the above discussion shows there is a difference between what happens at this scale for $T$- and $S$-type axions. For $T$-type axions the coupling to matter is order $1/M_p$ and this remains true above the KK scale. The breakdown of the 4D EFT is about the appearance of a multitude of new KK modes, all of which couple with gravitational strength. But the $S$-type axion's coupling to matter is proportional to $E/f_a$ and so actually grows to become order unity at the KK scale; what does this order unity coupling match to in the UV theory? It matches to a dimensionless extra-dimensional coupling in the UV theory: either to the coupling $c$ appearing in \pref{APPawedgeint} or to the coupling $g_{cs}$ of $B_{\MN}$ to the Chern-Simons term $S_{\MNP}$ that is implied by the field strength $G = \exd B + g_{cs} S$. Although  $g_{cs}$ is order $1/M^2$ when $B_{\MN}$ is dimensionless (as above), it is dimensionless once $B_{\MN}$ is canonically normalized in six dimensions. 

Since gauge invariance prevents the coupling \pref{APPbwedgeint} from containing a coupling between $T$-type axions and a gauge sector localized on a 3-brane, one can ask whether such couplings are more generally forbidden. The answer to this is `no' if we allow ourselves to consider gauge sectors localized on higher-dimensional branes. For instance for a six-dimensional 5-brane $\Sigma_6$ they can arise from an interaction of the form
\be
   S_{\rm int,g} = cM^2 \int_{\Sigma_6} B \wedge F \wedge F \propto c \int \exd^4 x \sqrt{-g} \; \mfb \, \epsilon^{\mu\nu\lambda\rho} F_{\mu\nu} F_{\lambda\rho}  \,,
\ee
where the explicit factor of $\cV_2$ coming from the integration over the additional two dimensions cancels the normalization of the harmonic form $\omega_{mn} \propto \cV_2^{-1} \varepsilon_{mn}$. For a canonically normalized scalar this would imply $g_{a\gamma\gamma} \sim1/ f_b$.

The upshot is this: the model-dependent $T$-type axions can couple surprisingly weakly to non-gauge matter compared to the scale set by their decay constant: $1/\cF \sim 1/M_p \ll 1/f_b \sim R_\KK$. By contrast, the model-independent $S$-type axion couples to matter with strength $1/\cF \sim 1/f_a \sim 1/f_b$ and the same is true of $T$-type couplings to gauge fields on higher-dimensional branes. From the point of view of the underlying string coupling $e^\phi$ the duality that maps $b_{\mu\nu}$ to $\mfa$ is also a strong/weak coupling duality.

\section{Conclusions}

Axions (or ALPs) are often motivated by appealing to string theory, which seems to provide them with abundance. But string theory also provides strong concrete evidence for the assertion that exact global symmetries cannot survive contact with quantum gravity; the observation that underlies the UV quality problem for attempts to solve the strong-CP problem using a global PQ symmetry.  

We here reconsider some of the implications that follow from the observation that axions arise as antisymmetric tensor fields in higher dimensions and that Standard Model fields usually live in localised objects like D-branes within the extra dimensions. Axions arise in two general types in this way: the model independent $S$-type axion originating from a two-form field in 4-dimensions after compactification; and the model dependent $T$-type axion such as arises as a Kaluza-Klein mode for an extra-dimensional tensor field (of which we focus for simplicity on two-form potentials in two extra dimensions).\footnote{Axions may come from other forms such as three or four forms in ten dimensions depending of which type of string theory.}

These allow for a rich structure of axion phenomenology and each type of axion can be adapted to realize the PQ solution to the strong CP problem. It has been known for a while that UV effects can affect the original PQ proposal by generating effective interactions that violate the global PQ symmetry and modify the prediction for the axion mass: the axion quality problem. We revisit how this problem arises for the two types of axion using the UV tools at hand.

We find that for $T$-type axions the quality problem resembles the form originally studied, since the UV theory directly provides a pseudoscalar field once compactified to four dimensions. Our framework differs from early versions of the quality problem that imagine the PQ symmetry to be linearly realized by a complex scalar at energies $E > f$, but generally agree with estimates based on the contributions due to wormholes or gravitational instantons below a compactification scale. To the extent that these contributions are exponentially suppressed their constraints are mild.

The $S$-type case is more interesting since both the strong CP and axion quality problems must first be reformulated in terms of the two-form field and its field strength. The PQ mechanism involves giving a mass to the axion and so on the dual side involves the `eating' of a 3-form potential along the lines proposed in \cite{Quevedo:1996uu}. The required 3-form potential is generated by the QCD sector itself as a non-propagating topological field in the EFT below the QCD scale. As applied to QCD our re-analysis broadly agrees with that of \cite{DualStrongCP} in concluding that the quality problem gets recast as an issue that arises when there are multiple 3-form fields present in the low-energy theory. This might be imagined to be a problem for string theory, for which 3-form potentials are as ubiquitous as axions. 

Prompted by recent discussions of this problem \cite{Sakhelashvili:2021eid, Dvali:2022fdv} we formulate the two properties which new 3-form fields must have if they are to threaten the PQ solution to the strong CP problem, arguing why string-generated 3-form fields are not generically a problem, largely because these fields need not couple to QCD (in string theory it depends on how bulk fields couple to brane fields and usually only one couples to QCD). The gravitational Chern-Simons term does couple to QCD but whether or not it sparks a new strong CP problem depends on whether or not gravity is strongly coupled in the UV. The discussions of \cite{DualStrongCP, Dvali:2022fdv} assume that it is, but we argue that if it is not (such as if the UV completion is a weakly coupled string vacuum) then the estimates for the size of the problem are again exponentially suppressed and so would not pose a quality problem.

Finally we also explore other UV implications for axion physics. We found that depending on the brane configuration hosting the Standard Model, extra dimensions can dramatically suppress physical couplings between the axion and Standard Model sector relative to the axion decay constant appearing in the axion kinetic term, especially if the volume of the extra dimensions is very large. This is possible for $T$-type axions but in the the examples examined does so only for non-gauge couplings (making this observation more pertinent for ALPs, whose properties would tend to be `fermiophobic'). 

For $S$-type fields both kinds of couplings have similar size.\footnote{For supersymmetric realizations this can be seen because both the K\"ahler potential and gauge kinetic function depend directly on the $S$ field.} For this case though, we argue that the duality relating the 2-form to the axion field swaps weak and strong couplings, and suggests a semiclassical description of 2-form response need not correspond to the usual semiclassical description of a scalar axion. This again motivates better exploring the 2-form side of the theory.

It is an old argument that UV information can have important implications for low-energy naturalness questions such as the strong CP problem. The observation that this could be informative in situations where the questions are solved using features like global symmetries that apparently should not be present at very high energies has sparked a revival of studies of generalised and non-invertible symmetries. Many of these ideas resonate well with string-motivated constructions, such as those we explore here.

\section*{Acknowledgements}
We thank Philippe Brax and Junwu Huang for helpful conversations. CB's research was partially supported by funds from the Natural Sciences and Engineering Research Council (NSERC) of Canada. Research at the Perimeter Institute is supported in part by the Government of Canada through NSERC and by the Province of Ontario through MRI.  
The work of FQ has been partially supported by STFC consolidated grants ST/P000681/1, ST/T000694/1.

\end{document}